\documentclass[ reprint,
 amsmath,amssymb,
 aps,
 prl
]{revtex4-2}
\usepackage{graphicx}
\usepackage{dcolumn}
\usepackage{bm}
\usepackage{physics}
\usepackage{xcolor}
\usepackage{subcaption}
\usepackage{float}
\usepackage[colorlinks=true,allcolors=blue]{hyperref}

\begin{document}

\title{Search for Axions and Dark Photons Using Single Molecule Magnets}

\author{Jose R. Alves$^{1,2,3}$, Manfred Lindner$^3$, Farinaldo S. Queiroz$^{1,2,4,5}$, Manoel S. Vasconcelos$^{1,6}$}
\affiliation{$^1$Departamento de F\'{\i}sica, Universidade Federal do Rio Grande do Norte, 59078-970, Natal, RN, Brasil
\\ $^2$International Institute of Physics, Universidade Federal do Rio Grande do Norte, Campus  Universit\'ario,  Lagoa  Nova,  Natal-RN  59078-970,  Brazil\\
$^3$Max Planck Institut fur Kernphysik, Heidelberg, Germany\\
$^4$ Millennium Institute for Subatomic Physics at the High-Energy Frontier (SAPHIR) of ANID, Fernandez Concha 700, Santiago, Chile\\
$^5$ Departamento de F\'isica, Facultad de Ciencias, Universidad de La Serena, Avenida Cisternas 1200, La Serena, Chile\\
$^6$Programa de P\'os Graduação em F\'{\i}sica, Universidade do Estado do Rio Grande do Norte,  Mossor\'o-RN, 59610-210, Brasil}

\date{\today}


\begin{abstract}
Molecular magnets, although analogous to familiar macroscopic magnets, offer a platform for next generation magnetic storage technologies with far higher data densities and prospective applications in quantum information science. When exposed to an external magnetic field, single molecule magnets enter a frustrated magnetic configuration that is exceptionally sensitive to low energy excitations. Energy deposited by a dark matter particle can trigger the relaxation of a metastable molecule, releasing Zeeman energy that subsequently propagates through neighboring molecules. This magnetic avalanche encodes the energy deposited in the initial excitation. By combining concepts from chemistry, condensed matter physics, and particle physics, we show that dysprosium and manganese molecules can achieve more than an order of magnitude improvement in sensitivity to dark photon and QCD axion models, respectively, compared with existing detection methods.
\end{abstract}

\maketitle

\section{\label{Intro}Introduction}

The existence of non-baryonic dark matter (DM), a dominant component shaping galactic structures and galaxy clusters, is firmly established through extensive cosmological and astrophysical evidence \cite{Cirelli:2024ssz}. Yet its fundamental nature remains elusive. The two leading interpretations—primordial black holes \cite{Green:2024bam} and elementary particles—present distinct pathways; here we focus on the latter. This approach necessitates non-negligible interactions between DM and Standard Model (SM) particles to enable experimental detection.

\begin{figure}[!h]
    \centering
    \includegraphics[width=0.9\linewidth]{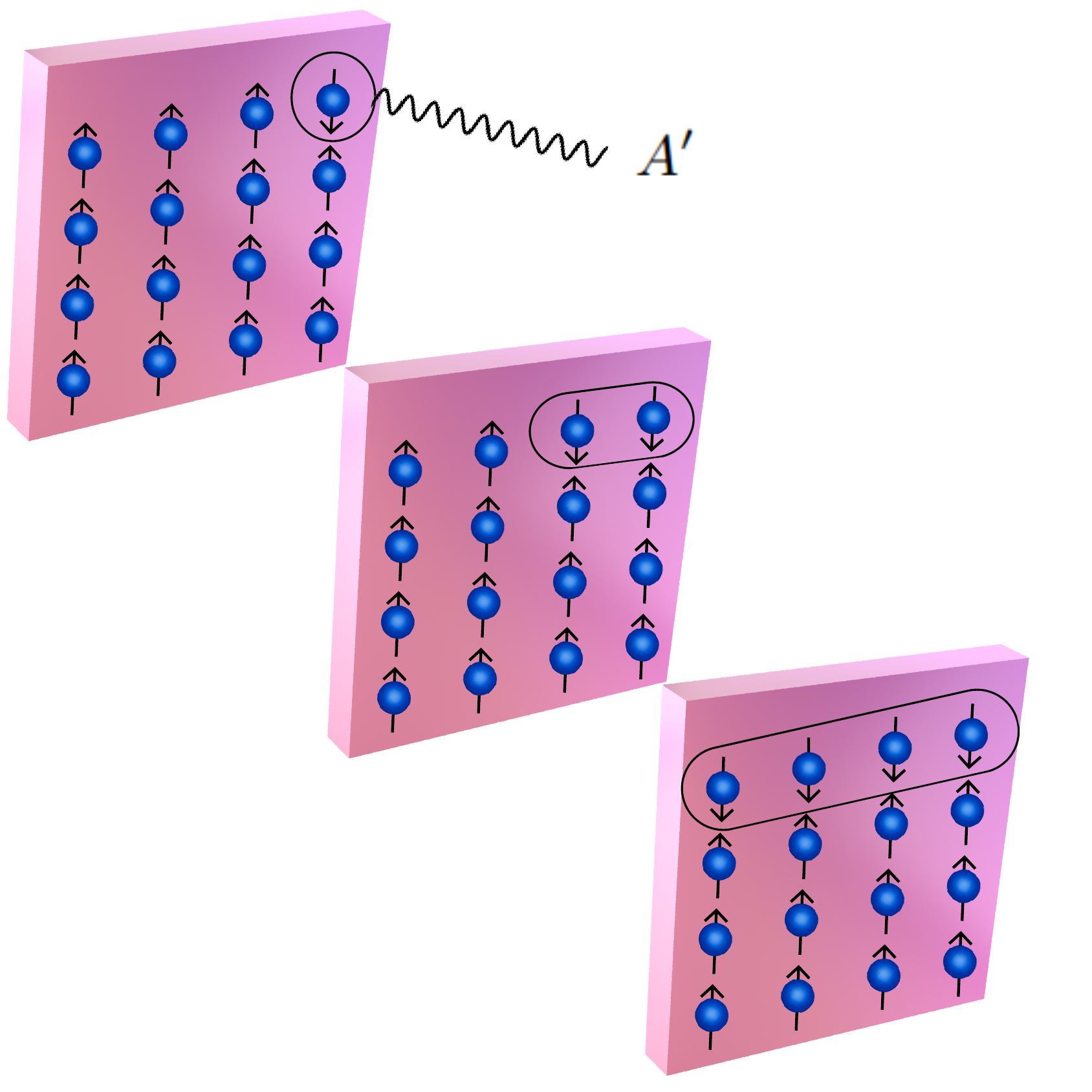}
    \caption{Sketch of the detection mechanism proposed in the paper. From left to right: the DM particle hits one of the molecules in the metastable state, which relaxes, releasing the Zeeman energy stored, which then propagates to the neighboring molecules, as shown by the red dashed circle increasing in size, in a process known as magnetic avalanche.}
    \label{representation-smm}
\end{figure}

Searches for DM via nuclear scattering \cite{Schumann:2019eaa, XENON:2023cxc, XENON:2024znc}, astrophysical signatures \cite{Gaskins:2016cha, Fermi-LAT:2016uux}, and accelerator production \cite{Boveia:2018yeb, PerezAdan:2023rsl} have excluded many minimal weakly interacting DM models. Future experiments like LUX-ZEPLIN \cite{Wang:2025ztb}, XENONnT, and PandaX-4T may soon achieve detection. However, for sub-GeV DM masses, nuclear recoils are suppressed.
A DM particle with mass $M$ will carry energy $E=1/2 M v^2 \sim  10^{-6}M$, yielding recoils below the keV thresholds of conventional detectors. While electron-recoil experiments extend sensitivity to MeV masses \cite{Essig:2012yx,Essig:2017kqs}, lighter DM remains inaccessible—motivating our novel approach.

Notably, dark matter particles in the Milky Way halo may undergo acceleration through elastic collisions with high-energy galactic cosmic rays and astrophysical neutrinos \cite{Ghosh:2024dqw,Agashe:2014yua,Bringmann:2018cvk,Das:2025qxm,Herbermann:2024kcy,Das:2024ghw}. This inevitable "boosted dark matter" (BDM) component, though small, carries sufficient kinetic energy to be constrained by direct detection experiments. However, when the mediator of the DM-target interaction is light, ($M_{med}\ll q$), sensitivity degrades due to momentum-dependent form factors. While existing BDM studies focus primarily on keV-scale masses, we target the unexplored sub-eV regime, where conventional detectors lack sensitivity.

\begin{figure}[h]
    \centering
    \includegraphics[width=0.9\linewidth]{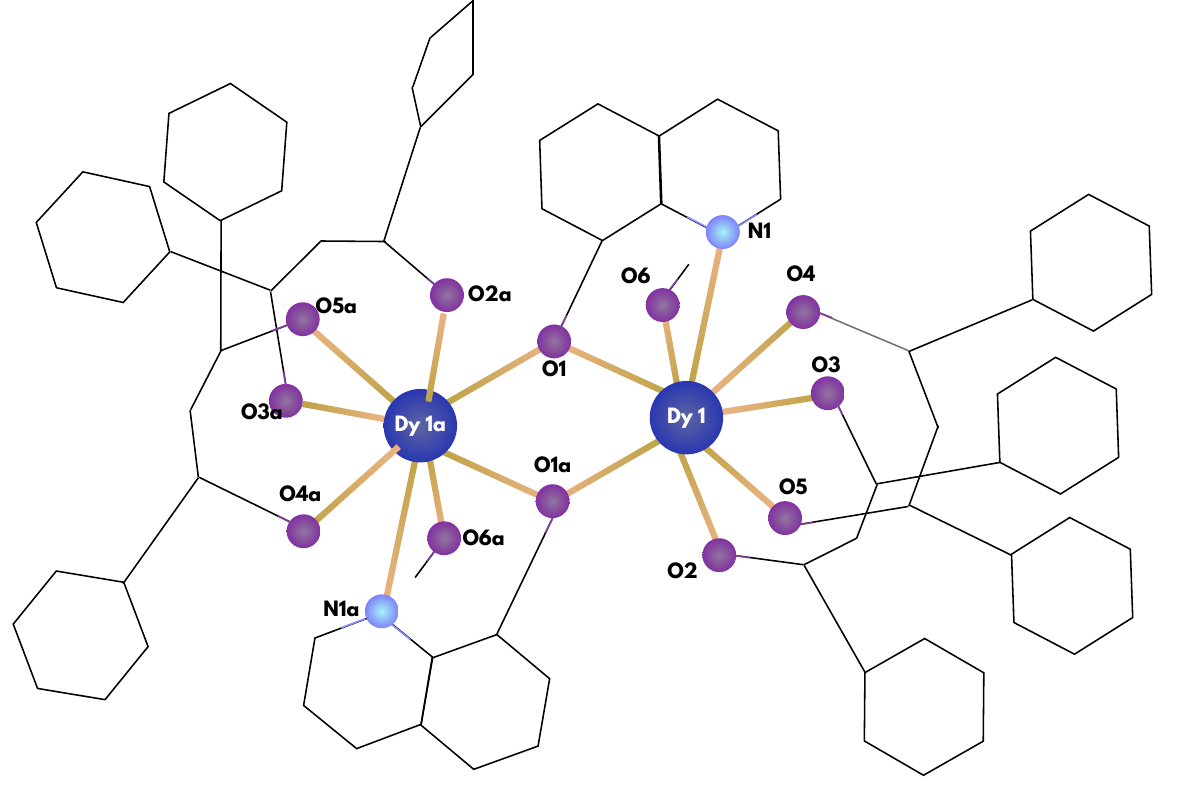}
    \caption{It illustrates the dinuclear dysprosium. The dark blue spheres represent the dysprosium nuclei, the purple circles the oxygen atoms, and the light blue spheres the nitrogen.}
    \label{fig:dinucleardisprosium}
\end{figure}

\begin{figure}[h]
    \centering
    \includegraphics[width=0.9\linewidth]{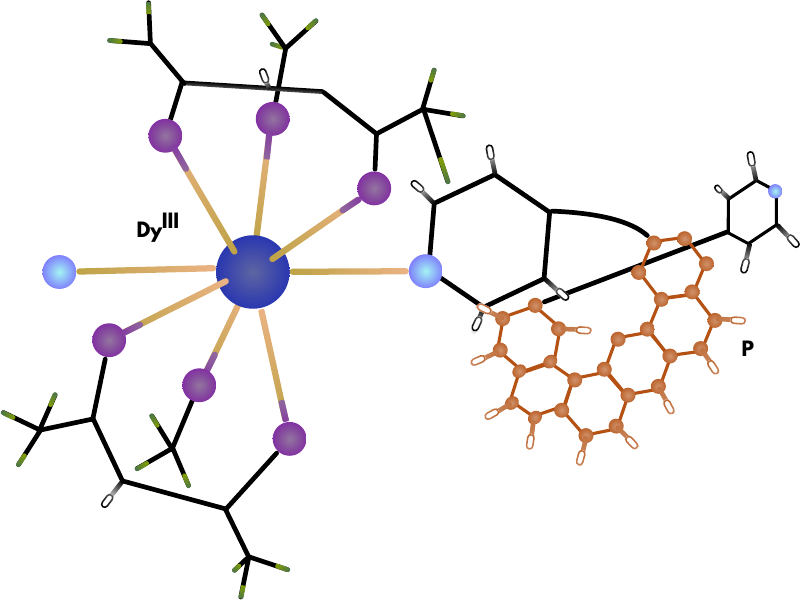}
    \caption{Figure displaying Dy$^{\text{III}}$. The dysprosium atom is in dark blue, oxygen in purple, and nitrogen in light blue. Highlighted in orange is the helical helicity center for the compound, which drives the compound's chiral features.}
    \label{fig:magnetochiralimage}
\end{figure}

Motivated by this gap, we leverage condensed-matter systems—building on proposals using semiconductors \cite{Hochberg:2016sqx,Kahn:2021ttr}, photonic nanostructures \cite{Herrera:2016hlp}, Fermi-degenerate materials \cite{Hochberg:2015fth}, and magnetic bubbles \cite{Bunting_2017}. Specifically, we propose dysprosium-based single-molecule magnets (SMMs) as BDM detectors: the dinuclear complex Dy$_2$, and the complex Dy$^{\text{III}}$ (illustrated in FIGs.\ref{fig:dinucleardisprosium}-\ref{fig:magnetochiralimage}) both exhibiting robust SMM behavior \cite{magneto-chiral, near-infrared}.
We further incorporate far-infrared spectra of Manganese, $Mn_{12}$-acetate \cite{mn12data1, mn12data2} to probe {\it axion} masses across complementary frequency bands.

These systems maintain metastable magnetic states (See. FIGs.\ref{drawing1}-\ref{drawing2}) for extended periods at cryogenic temperatures 
 $T\sim 0.1$~K. They have attracted considerable interest due to their applications in magnetic storage and potential use as qubits in quantum computing \cite{Friedman_2010}. This characteristic of staying in a false vacuum with a relatively long period allows us to probe dark sectors via the absorption of a BDM as illustrated in  FIG.\ref{representation-smm}.

We derive the physics reach of this new method for two popular dark matter models: {\it dark photon} and {\it axion}. The {\it dark photon} arises as the gauge boson of an extended $U(1)'$ symmetry in the SM Lagrangian via the interaction term, \cite{Fabbrichesi_2021},
\begin{equation}
    \mathcal{L}_{\text{DP}} \supset   - \frac{\varepsilon}{2}F_{\mu \nu}F'^{\mu \nu},
    \label{darkphoton-lagrangian}
\end{equation}where $\epsilon$ is the kinetic mixing. The {\it axion} particle has been introduced to solve the strong CP problem as a result of the Peccei–Quinn symmetry \cite{Peccei:1977hh,Preskill:1982cy,Kim:2008hd}. In the end, this solution induces a coupling to the electromagnetic field strength tensor described by \cite{axion-absorption, ParticleDataGroup:2024cfk},
\begin{equation}
    \mathcal{L}_{\text{a}} \supset -\frac{1}{4}g_{a\gamma\gamma} a F_{\mu \nu} \tilde{F}^{\mu \nu}.
    \label{eq:axionphotoncoupling}
\end{equation}

Because this work spans several disciplines, we begin by introducing the essential concepts of SSMs and molecular absorption. We then show that, for particle masses below 1 eV, these systems provide a uniquely powerful platform where particle physics, condensed matter physics, and chemistry intersect, enabling a detection strategy with the potential to surpass the sensitivity of existing experimental approaches by more than one order of magnitude.
\noindent   

\section{Single Molecule Magnets
\label{SMM}}
In this section, we outline the key properties of SSMs that underpin their use as dark‑matter sensors and review the essential principles governing absorption in SSMs to establish how the ensuing magnetic avalanche encodes the energy deposited by an incident DM particle.

\begin{figure}[h]
    \centering
    \begin{subfigure}[b]{0.49\textwidth}
        \centering
        \includegraphics[width=0.6\textwidth]{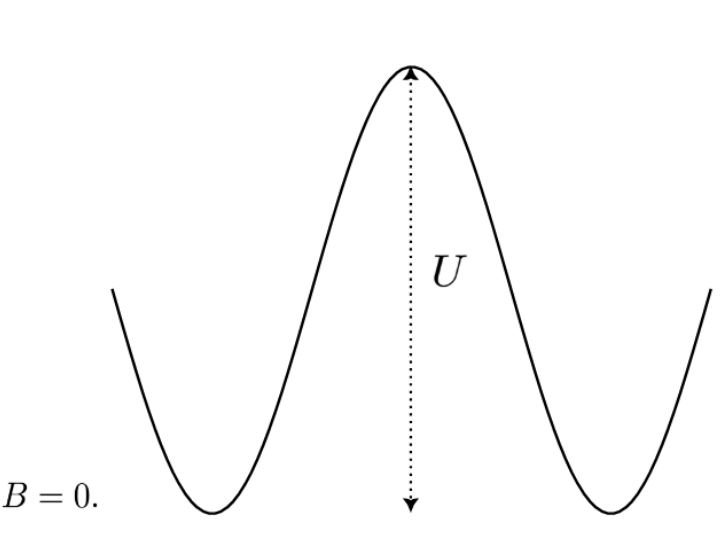}
        \caption{Splitting of the potential felt by an SMM without an external magnetic field.}
        \label{drawing1}
    \end{subfigure}
    \hfill
    \begin{subfigure}[b]{0.49\textwidth}
        \centering
        \includegraphics[width=0.6\textwidth]{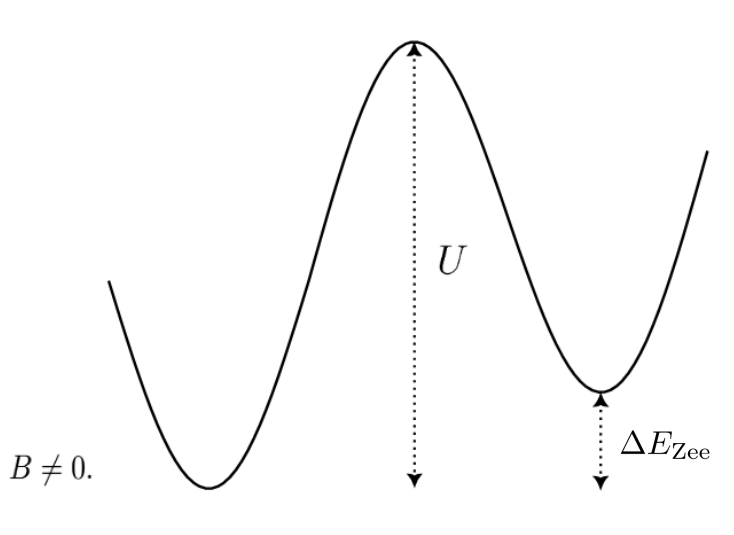}
        \caption{Splitting of the potential felt by an SMM with an external magnetic field.}
        \label{drawing2}
    \end{subfigure}
    \caption{The potential felt by an SMM molecule, FIG.\ref{drawing1} shows the potential without an external magnetic field, picturing the degenerate vacuum, while FIG.\ref{drawing2} pictures one of the vacuum lifted due to the presence of an external magnetic field. The mechanism proposed focuses on the latter.}
    \label{drawings}
\end{figure}

\begin{figure}[h]
    \centering
    \includegraphics[width=1\linewidth]{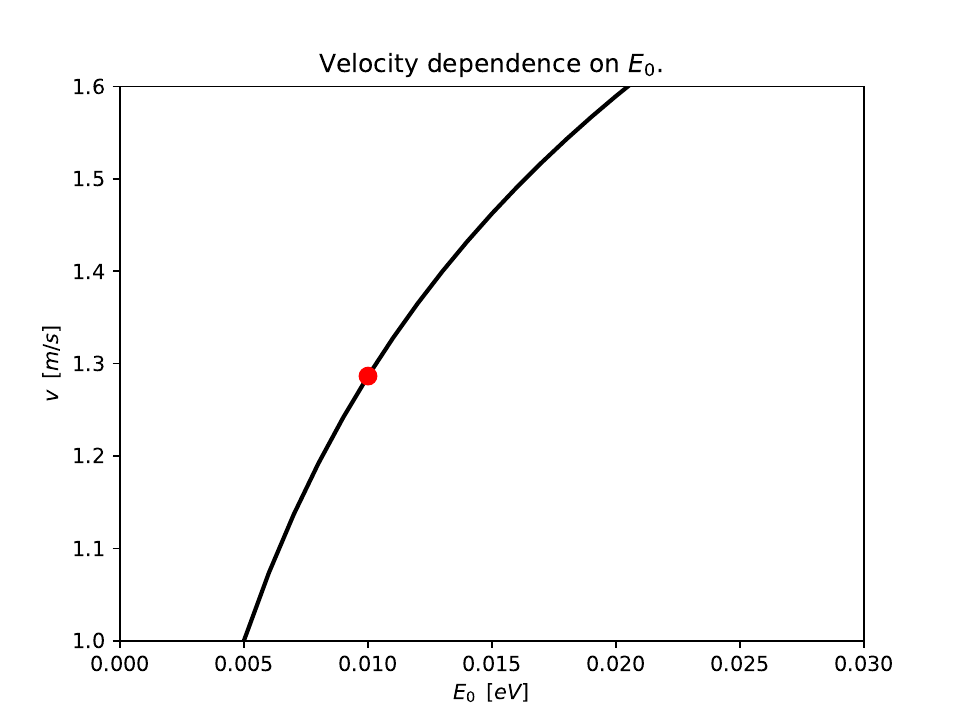}
    \caption{Velocity dependence on the deposited energy based on Eq. \eqref{eq-velocity}. The behavior is as expected: the greater the energy deposited, the higher the velocity. The red dot shows the expected velocity for an energy deposition around $10^{-2}$eV. This plot is calculated for a Zeeman energy of $10^{-4}$eV for an anisotropic energy barrier of $U = 100$ K. The velocity also depends on how much Zeeman energy is stored, which means that the general behavior will change by changing the temperature at which the sample is kept.}
    \label{velocitydependence}
\end{figure}

SMMs exhibit magnetic properties similar to those of bulk magnetic material. The term {\it single-molecule} magnet means that the considered molecule is a magnet itself, with a unique domain. In bulk magnets, the magnetic moment arises from the presence of unpaired electrons, meaning their spins form parallel/antiparallel arrangements within domains (this is what we call the areas separated by Bloch domain walls). As well as in bulk materials, the SMMs can be magnetized by an external magnetic field, and are demagnetized when this external magnetic field is removed. We work with lanthanydes, which are a class of f-block metals, in order to describe its magnetic properties, we need to consider spin–orbit coupling terms, characterized by the quantum number J, i.e., the total angular moment, and their splitting by the applied field, leading to sublevels that are described by $E_J$ instead of quantum numbers $E_S$ \cite{ZABALALEKUONA2021213984}.

It has been found that its magnetization slowly relaxes with the removal of an applied external magnetic field, and the relaxation timescale is different from that of bulk materials; however, some molecules can have their relaxation timescale made large enough for the proposed experiment for low temperatures, as in Eq. \eqref{eq:orbach}. The relaxation time for this process will be discussed below. Therefore, these new molecules have induced a new concept in the field of magnets, generating interesting experimental and theoretical consequences, such as out-of-phase signals in AC studies of magnetic properties \cite{Orts}. Most of these molecules exhibit a large effective spin ground state $S$ that may appear due to exchange of ions in the magnetic core (as is the case for Mn12-acetate, whose S is equal to 10 \cite{Sessoli,ZABALALEKUONA2021213984}) or competing antiferromagnetic interaction between the spins of the ions, alongside an axial zero-field splitting which separates the $(2S + 1)$ $m_s$ projections of the spin in the $z$-axis (or easy axis of the molecule). This splitting is characterized by a parameter $D$ present in the Hamiltonian that describes these molecules, which is given by,
\begin{equation}
    H \supset -D S_z^2.
    \label{hamiltonian-anistropy}
\end{equation}
This Hamiltonian nicely describes the system at low temperatures. For $D > 0$  the low-lying energy levels are those with the highest $|E_S|$ value, in this case $E_S = \pm S$. The magnetization of the molecule is associated with each of the $E_S = \pm S$ sublevels having a particular orientation along the axial anisotropy axis; thus, $E_S = +S$ corresponds to spin up, whereas $E_S = -S$ corresponds to spin down.
A pictorial view of the splitting with and without an applied external magnetic field can be found in FIG.\ref{drawings}.

In FIG.\ref{drawing1}, we plot the crossover between the energy levels and a double well, including a barrier of size $U$, indicating on the left the negative $E_S$ levels and on the right side, the positive $E_S$, just to illustrate the degeneracy of the eigenstates, without an applied external magnetic field. In FIG.\ref{drawing2}, we exhibit the potential in the presence of an external magnetic field, showing a metastable vacuum state with energy equal to $2\mu_B\, g\, J\, B$ with respect to the true vacuum, where $\mu_B$ is the Borh magnetic moment, g is the gyromagnetic factor, J the total angular momentum and B the magnitude of the applied magnetic field. 

\begin{figure}[h]
    \centering
    \includegraphics[width=1\linewidth]{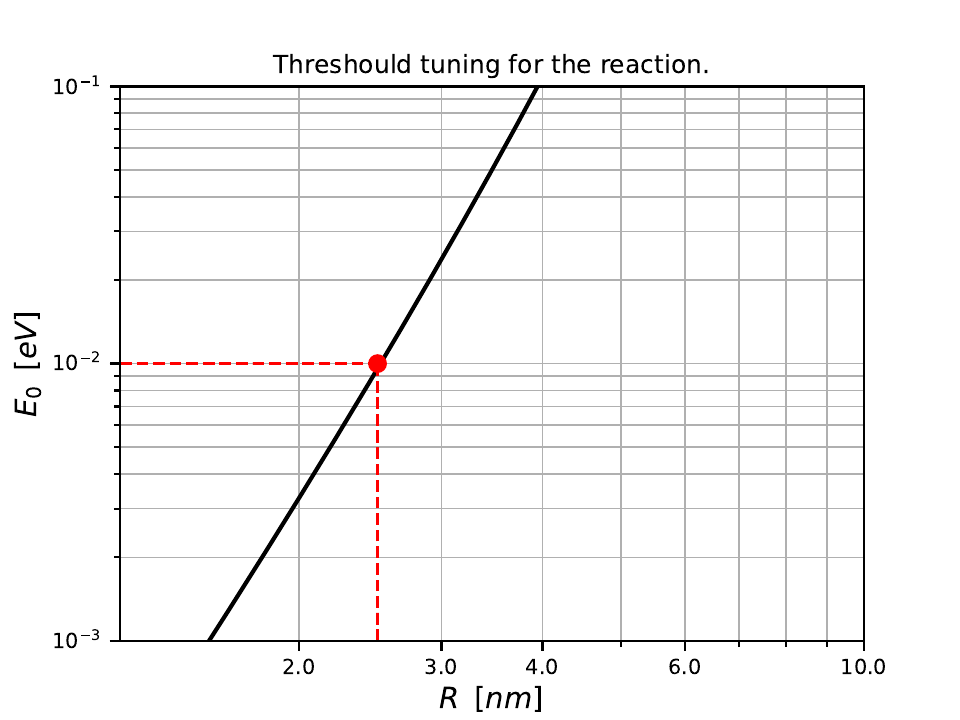}
    \caption{Tuning of the necessary deposited energy to trigger the reaction. As can be seen for a radius of $R \approx 3$nm, which is around the value for a SMM of Dy$^{\text{III}}$, the energy deposition is of at least $10^{-2}$eV as shown by the red dot and lines.}  
    \label{tuning}
\end{figure}

These nanomagnets can go from one state to another via various relaxation mechanisms \cite{C9SC01062A, ZABALALEKUONA2021213984,abragam2012electron}. The most common relaxation processes operating in SMMs are the ones that occur through spin-phonon coupling (phonon being the quantum of the vibration mode of the molecules), namely Orbach, Raman and direct processes. Also, we have those that occur due to the quantum nature
of the materials, generally quantum tunnelling of the magnetization
and thermally assisted relaxation processes \cite{abragam2012electron}. For our case, we are working with lanthanides, f-block metals \cite{ZABALALEKUONA2021213984} in which the important mechanisms are
Orbach and Raman \cite{Briganti2021}. These involve two-phonon-assisted processes, while the others can involve a single or two phonons. For the Orbach relaxation process, the system
overcomes the entire barrier $U$ (See FIG.\ref{drawing2}) by absorbing  phonons
from the crystal lattice containing the exact energy $\hbar \omega_i$ to jump from one of the ground sublevels ($E_S = \pm S$) to the highest excited sublevel ($E_S = 0$), with $\omega_i$ being the frequency of the incoming phonons, and so, from this excited state, the system will relax to either the state with $E_S =- S$ or the state with $E_S=+S$ emitting new phonons with energy $\hbar \omega_e$, $\omega_e$ being the frequency of the emitted phonon. Therefore, the energy difference between the absorbed and emitted phonon will correspond to the energy difference between the $\pm E_S$ sublevels
in the ground state. The Orbach process does not occur between the highest excited states. Instead, this process usually occurs between the first or second excited states. So, a large phonon energy is required for this process to occur, and it usually operates at the highest temperatures. However, for the Raman, the assisted process is driven by the inelastic dispersion of phonons. In this process, the molecule will absorb a phonon with energy $\hbar \omega_i$, reach a virtual excited state, and emit another phonon with energy $\hbar \omega_e$. In that way, the energy difference between the two phonons will correspond to the energy difference between the
$\pm E_S$ sublevels in the ground state. 

In fact, all these assisted processes happen due to spin-phonon interactions that require energy in the form of thermally assisted energy. For example, a recent study shows, through ab-initio calculation for a Dy${}^{\text{III}}$ based molecule \cite{Briganti2021}, that the two relaxation processes that dominate are: Orbach and Raman for high and low temperatures, respectively. As we are interested in the absorption mechanism that injects sufficiently large energies, we use the Orbach relaxation description (for a review, see \cite{ZABALALEKUONA2021213984}), 
\begin{equation}
    \tau_R \approx \tau_0\exp\left( \frac{U - \frac{1}{2}\Delta E_{\text{Zee}}}{\Delta T} \right),
    \label{eq:orbach}
\end{equation}
in which $U$ is the size of the energy barrier related directly to $D$ and $\Delta E_{\text{Zee}}$ is the stored Zeeman energy, $\tau_0$ is the attempt time related to each type of SMM \cite{suzuki_mn12}, and $\Delta T$ is the temperature difference.

Since these molecules might be in a metastable state as shown in FIG \ref{drawing2}, if an $O(1)$ fraction of SMMs is kept in this state, which can be achieved if the system is kept at sufficiently low temperatures ($\sim 1$ K), we can use them as a mean of directly detecting dark matter. The relaxation of them from the metastable state and thus the release of the stored Zeeman energy in the form of thermal energy induces the relaxation of neighbor molecules and so on, as shown in FIG.\ref{representation-smm} which sketches the deflagration mechanism, that is known as magnetic avalanche. This process is known and reported and is at the heart of the system's use as a detector and can be measured using micron-sized Hall sensors \cite{Friedman_2010, PropagationofAvalanches}.

\begin{figure}[h]
    \centering
    \begin{subfigure}[b]{\linewidth}
        \centering
        \includegraphics[width=\linewidth]{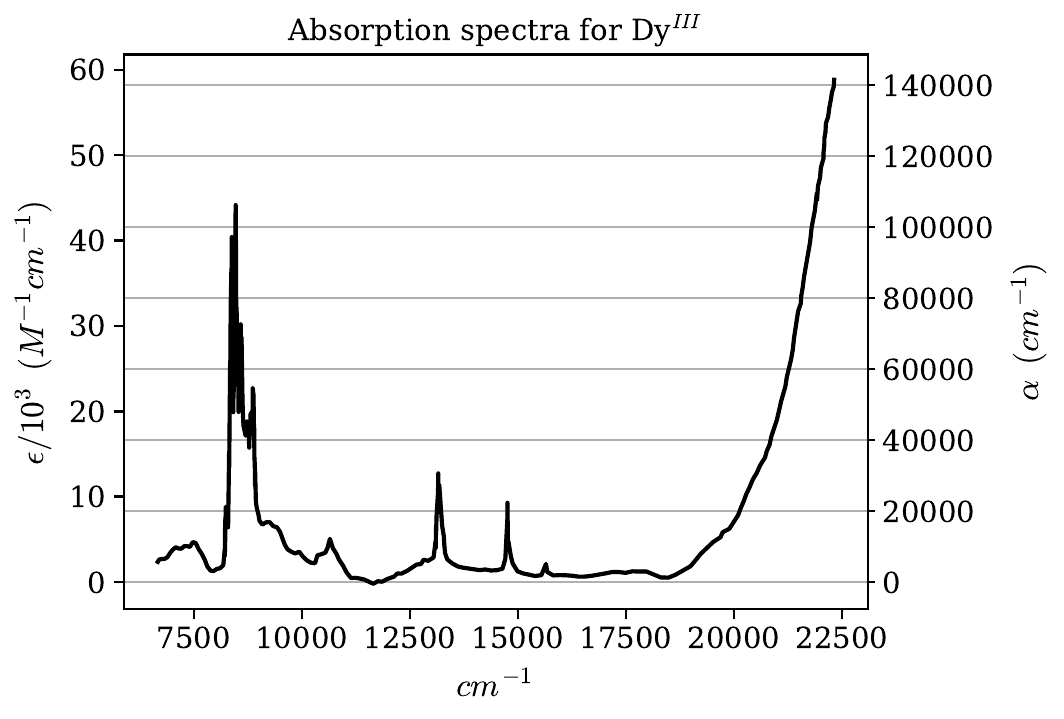}
        \caption{}
        \label{magneto-chiralimage}
    \end{subfigure}
    \hfill
    \begin{subfigure}[b]{\linewidth}
        \centering
        \includegraphics[width=1\linewidth]{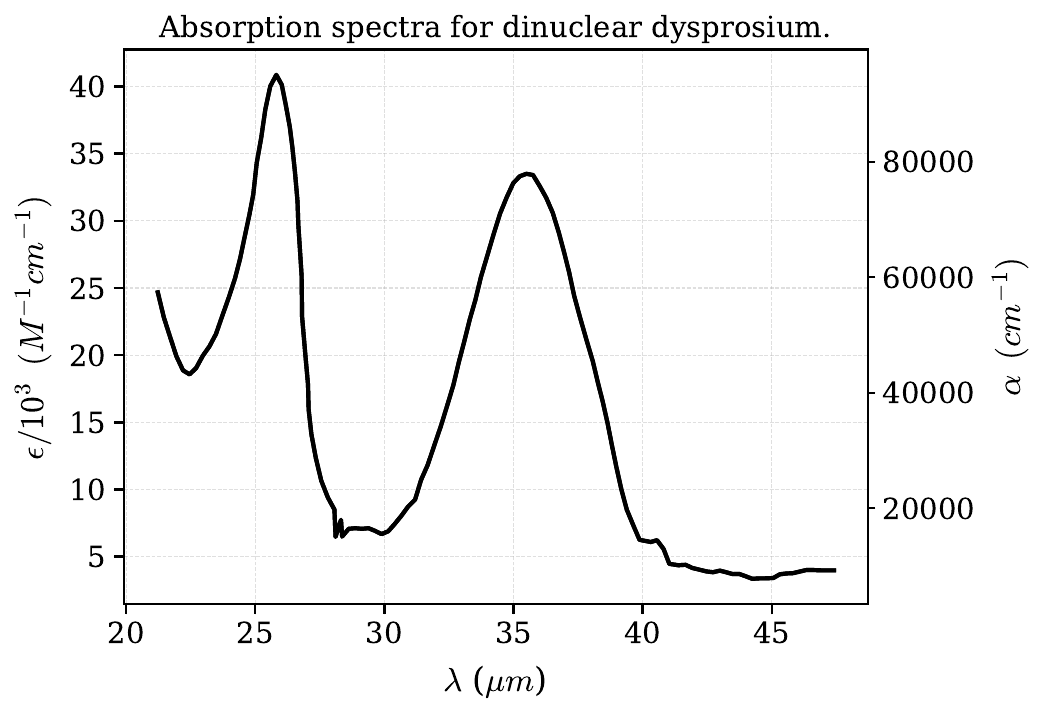}
        \caption{}
        \label{absorptionDY2}
    \end{subfigure}
    \caption{Absorption spectra of both molecules are considered in the analysis. FIG.\ref{magneto-chiralimage} is the solid-state absorption spectra of UV-vis-NIR for the crystal \textbf{1-(P)} as reported from \cite{magneto-chiral}. The x-axis is the wavenumber in $cm^{-1}$ and the y-axis is the absorptivity in $cm^{-1}$. FIG.\ref{absorptionDY2} is the absorption spectra of UV-vis light as reported from \cite{near-infrared}. This is the curve that corresponds to complex Dy$_2$(dbm)$_4$(OQ)$_2$(CH$_3$OH)$_2$ synthesized by 8-hydroxyquinoline and dibenzoylmethanate (dbm) ligands. The x-axis corresponds to the wavelength of light, and the left y-axis corresponds to the molecular absorption coefficient and the right y-axis to the absorption coefficient that directly relates to the cross-section via Eq. \eqref{eq:abs-crossection}.}
    \label{absorptions}
\end{figure}

\begin{figure}[h]
    \centering
    \includegraphics[width=0.9\linewidth]{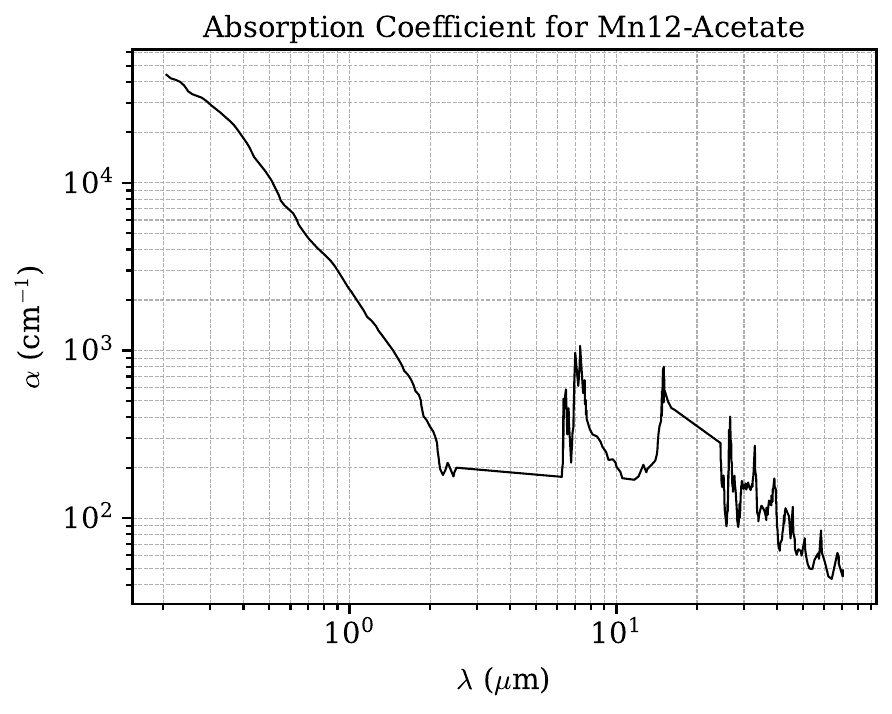}
    \caption{Absorption coefficient for the Mn12-Acetate, which were taken from \cite{mn12data1, mn12data2}, considered in the calculation of the kinetic mixing of \cite{Bunting_2017} and on the {\it axion}-photon coupling of our work. The flat region in the wavelength interval of $10^{0} < \lambda < 10^1$ is an interpolation due to the lack of data.}
    \label{fig:mn12absorption}
\end{figure}

Magnetic avalanche is a process in which the rapid release of magnetic energy occurs along a given neighbourhood, in a similar way to a cascade. This process usually occurs in solar flares \cite{2001SoPh}, molecular magnetic materials\cite{PropagationofAvalanches} and in some types of diodes \cite{McLuckie_1988}. For molecular crystals, it is possible to demonstrate this phenomenon through local measurements with temporal resolution of the rapid magnetization reversal in millimeter-sized single crystals of Mn12 acetate \cite{PropagationofAvalanches}. In this case, the magnetic avalanche takes the form of a thin interface between regions of opposite magnetization. This interface propagates throughout the crystal at a constant speed for a fixed applied field. In fact, the propagation of this interface alters the local magnetization of the molecule within the crystal (see FIG.\ref{representation-smm}) in a similar way to a cascade. Another way to understand this phenomenon is that it is quite analogous to the propagation of a flame front (deflagration) through a flammable chemical substance.

The velocity of magnetic avalanches can be given by a crude estimate using the heat equation \cite{PropagationofAvalanches} for a sample of Mn12-acetate, and can be used to assess the sensitivity of the reaction to that of the deposited energy. The velocity can be determined using Hall sensors that measure the magnetic field in a direction perpendicular to the applied field. Our proposal relies on a different sample, but the same relaxation process drives the deflagration mechanism once DM hits the detector, that is the Orbach routine \cite{PropagationofAvalanches, Briganti2021}. Hence, we can use this velocity to assess the deposited energy of the DM particle,  
\begin{equation}
    v \sim \sqrt{\frac{\kappa}{\tau_0}}\exp\left(- \frac{U - \frac{1}{2}\Delta E_{\text{Zee}}}{\Delta T} \right),
    \label{eq-velocity}
\end{equation}
where $\kappa$ is the thermal diffusivity of the SMMs. This magnetic avalanche propagates as the narrow interface through the crystal at a constant velocity \cite{Friedman_2010, PropagationofAvalanches}. The temperature difference can be put in terms of the deposited energy $E_0$ from the DM particle (see Sec. \ref{tuning-section}), which can serve as a probe for the signal; this feature can be seen in FIG.\ref{velocitydependence}.

\noindent

\section{Setting up the Detector \label{tuning-section}}

In order to find the necessary energy to trigger the magnetic avalanche reaction in a SMM detector, we can follow \cite{Bunting_2017} in comparing the characteristic time of relaxation, $\tau_R$ for an Arrhenius law ($\tau_R \approx \tau_0 \exp{(\frac{U - \frac{1}{2}\Delta E_{\text{Zee}}}{\Delta T})}$, where $\tau_0$ is a constant related to each SMM and it may vary from $10^{-12}-10^{-9}$, $\Delta E_{\text{Zee}} = 2 \mu_B g_J J B$ is the Zeeman splitting, $U$ the size of the energy barrier and $\Delta T$ the difference in temperature) with the time heat takes to dissipate in the detector, $\tau_D$; considering that we need the SMM to flip before heat dissipates, we need $\tau_R < \tau_D$. With this imposition, we can sense the minimum energy needed if we assume that the SMM will have a Debye behavior for its temperature dependence, which is the observed behavior for SMMs at low temperatures \cite{Miyazaki2001}. This approximation yields the following,
\begin{equation}
    E_0 \gtrsim \frac{c_0 R^3(U- \frac{1}{2}\Delta E_{\text{Zee}})^4}{\ln{\left(\frac{R^2}{\tau_0 \kappa}\right)}},
    \label{energy-threshold}
\end{equation}where $c_0$ and $R$ are the SMM's volume-specific heat capacity and radius.

\begin{figure*}[t]
    \centering
    \includegraphics[width=\textwidth]{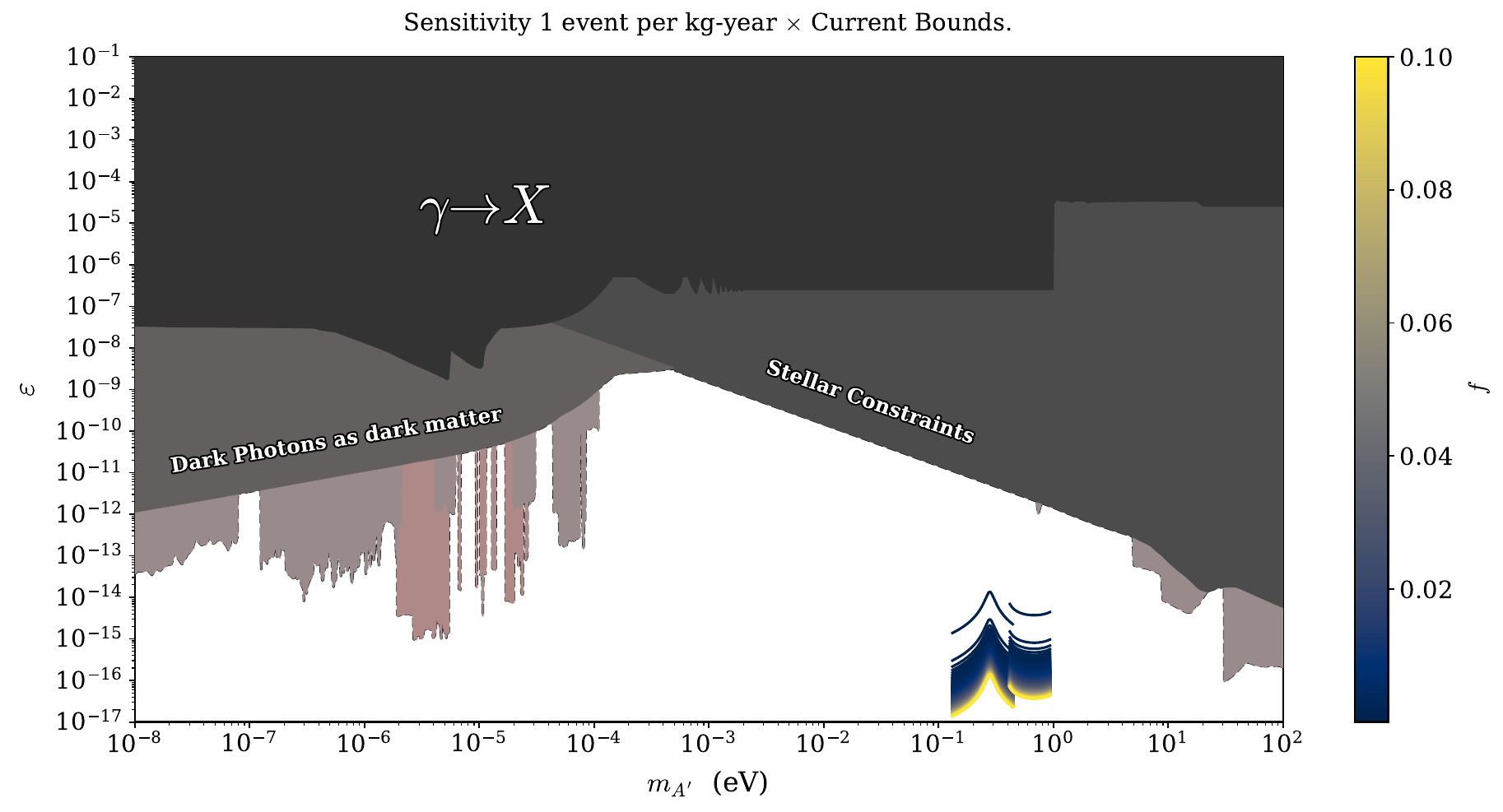}
    \caption{Expected sensitivity to the kinetic mixing, $\varepsilon$, which governs the interaction with SM fermions. We varied $f=0.01-0.1$. Clearly, dysprosium can be a promising laboratory for {\it dark photons} with the potential to cover a few orders of unexplored parameter space.}
    \label{fig:dykgyearbounds}
\end{figure*}

Eq. \eqref{energy-threshold} places a lower bound on the energy threshold as a function of $R$ for SMMs. Using the data reported in \cite{near-infrared} for the dinuclear Dysprosium illustrated in FIG.\ref{fig:dinucleardisprosium} that features $\tau \sim 4 \times 10^{-9}$s,$U = 109.5$K and $\kappa \sim 10^{-7} \text{m}^2/\text{s}$ at $1K$ \cite{Bunting_2017}, while for Dy$^{\text{III}}$ \cite{magneto-chiral} does not report the value of $\tau_0$, we adopted the same value.

The results are shown in FIG.\ref{tuning}, where we can see that for an SMM molecule of radius $\sim (2-3)$nm, the threshold energy to trigger the reaction will be $\sim 10^{-2}$eV. Given that the DM velocity in our galaxy is of order $v\sim 10^{-3}$, the average deposited energy is approximately $\sim 10^{-7} - 10^{-6}$eV, which is insufficient to trigger the reaction. To account for this, we consider that only a fraction $f$ of DM particles will have sufficient energy to initiate the reaction. These particles may originate from a class of physical processes that impart additional energy to DM, a scenario known as Boosted Dark Matter (BDM) \cite{Bringmann:2018cvk, Super-Kamiokande:2022ncz, Granelli:2022ysi, Das:2025qxm, Herbermann:2024kcy, Das:2024ghw}.

\section{Results \label{Results}}

In this section, we demonstrate the new physics reach of this proposed detection method for the {\it dark photon} and {\it axion} models. In particular, we will assess the sensitivity to the kinetic mixing parameter $\varepsilon$ for the {\it dark photon} dark matter (DPDM), $A_\mu'$, and the {\it axion}-photon coupling, $g_{a\gamma \gamma}$, for {\it axion} model. Starting with the {\it dark photon} model, it is well-known that the kinetic mixing term between the photon and the dark photon in Eq. \eqref{darkphoton-lagrangian} induces an interaction between with SM fermions described as follows,
\begin{equation}
    \mathcal{L}_{\text{DP}} \supset e\varepsilon A'_\mu J^\mu,
    \label{interaction-dp-sm}
\end{equation}
which is very similar to QED, but with the electric charge suppressed by the kinetic mixing angle $\varepsilon$. Consequently, the interaction cross-section of SM fermions and the dark sector gauge boson can be expressed as the QED cross-section rescaled by $\varepsilon^2$ \cite{Fabbrichesi_2021},
\begin{equation}
    \sigma_{A'} = \varepsilon^2 \sigma^\gamma.
    \label{darkphotoncrosssection}
\end{equation}

We can reinterpret the average absorption cross section for the {\it dark photon} model
\cite{Bunting_2017,An_2015},
\begin{equation}
    \langle n \sigma_{\text{abs}}(m_{A'})v\rangle \approx \varepsilon^2 \langle n \sigma^\gamma_{\text{abs}} (c = 1)\rangle.
    \label{crossectionrelations}
\end{equation}
This approximation arises because the mixing angle for absorption is rescaled by the relative permittivity of the SMM, which can be approximated as unity \cite{Bunting_2017}. This approximation, alongside the rescaling of the {\it dark photon} cross-section, is helpful because the actual cross-section for photons being absorbed in materials is known \cite{Bunting_2017},
\begin{equation}
    \sigma^\gamma_{\text{abs}} = \frac{\alpha}{n},
    \label{eq:abs-crossection}
\end{equation}
where $\alpha$ is the absorption coefficient of the material and $n$ its number density. While for the {\it axion}, its coupling to the photon is via a dimension five operator in Eq.\eqref{eq:axionphotoncoupling}, which leads to \cite{ParticleDataGroup:2024cfk},
\begin{equation}
    \mathcal{L}_{\text{a}} = g_{a \gamma \gamma}\, a \,\vec{E} \cdot \vec{B},
\end{equation}
where $g_{a\gamma \gamma}$ has dimension of Energy${}^{-1}$. We can determine the total rate of absorption of an {\it axion} in a magnetized medium per unit exposure,
\begin{equation}
    R \simeq \left( \frac{g_{a \gamma \gamma} B_0}{m_a} \right) \frac{\rho_{\text{DM}}}{\rho_\text{T}} \text{Im} \left[ -\frac{1}{\kappa(m_a)} \right],
\end{equation}
where $m_a$ is the {\it axion}'s mass, $\rho_\text{DM}$ and $\rho_\text{T}$ are the DM local density, and the target's density, $B_0$ is the applied magnetic field in the medium, and $\kappa(m_a)$ is the dielectric function evaluated at the {\it axion}'s mass \cite{axion-absorption, Hochberg_2016, Mitridate:2021ctr, Catena:2022fnk} and,
\begin{equation}
    g_{a \gamma \gamma} \sim \frac{\varepsilon m_a}{B_0}.
    \label{eq:axion-dpdm}
\end{equation}

Via Eq. \eqref{eq:axion-dpdm}, we translate the expected sensitivity to the {\it axion}-photon coupling. Eq. \eqref{eq:axion-dpdm} also shows that the {\it axion}-photon coupling is amenable to the strength of the applied magnetic field, contrary to the kinetic mixing sensitivity, which has no dependence at all in our case. This is a considerable feature for the {\it axion}, because we can improve the sensitivity through strong magnetic fields, while reducing the energy threshold in Eq. \eqref{energy-threshold}. Thus, if we lower the threshold enough, we might not need to work in the framework of Boosted Dark Matter anymore. Nonetheless, to be consistent with the {\it dark photon} case, we will consider both with the same value of applied magnetic field: $B = 1$T. Therefore, there is still a need for the BDM framework as Eq. \eqref{energy-threshold} is model independent.

We need to derive the absorption coefficient, $\alpha$, for the molecules in question to obtain the absorption rate. To do so, we need to use absorbance or transmittance via Beer-Lambert's law,
\begin{equation}
    A = -\log_{10} e^{-\alpha hl},
    \label{beer-lambert1}
\end{equation}
where $h$ is the concentration and $l$ the thickness of the material under analysis. To obtain data for the molecules considered, we used \cite{near-infrared, magneto-chiral}. However, the reported data is in terms of the molecular absorption coefficient as can be seen in the left y-axis of FIG. \ref{absorptions}, which relates to Beer-Lambert's law in a different form,
\begin{equation}
    A = \epsilon \times c \times l,
\end{equation}
where $\epsilon$ is the molecular absorption coefficient, $c$ is the concentration given in mol$/L = M$, which makes $[\epsilon] =\text{cm}^{-1} \text{M}^{-1}.$ The conversion from one to the other is made considering that the molar concentration can be related to the concentration in Eq. \eqref{beer-lambert1} by $c = \frac{\rho\times h}{m}\times \frac{10^3}{N_A}$, for $\rho, m$, and $N_A$ are the density, mass and Avogadro's number of the molecule. Using the molar mass of the substance in consideration, $\mathcal{M} = \frac{m}{N_A}$, we get the exact relationship between both coefficients,
\begin{equation}
    \alpha = 10^3\ln{10} \frac{\rho}{\mathcal{M}}\epsilon.
\end{equation}

Using this equation and the data reported in \cite{near-infrared, magneto-chiral} for the density and molecular mass of the substances, we get the exact values for $\alpha$ as a function of wavenumber. The values are shown in the y-axis of FIG.\ref{absorptions}, where the shape for $\epsilon$ and $\alpha$ are the same, because the conversion is independent of the wavelength under consideration. With the exact values for $\alpha$ we can then directly connect the reported data with the prospects for the sensitivity of a DPDM particle being absorbed in the detector.

\begin{figure*}[t]
    \centering
    \includegraphics[width=\textwidth]{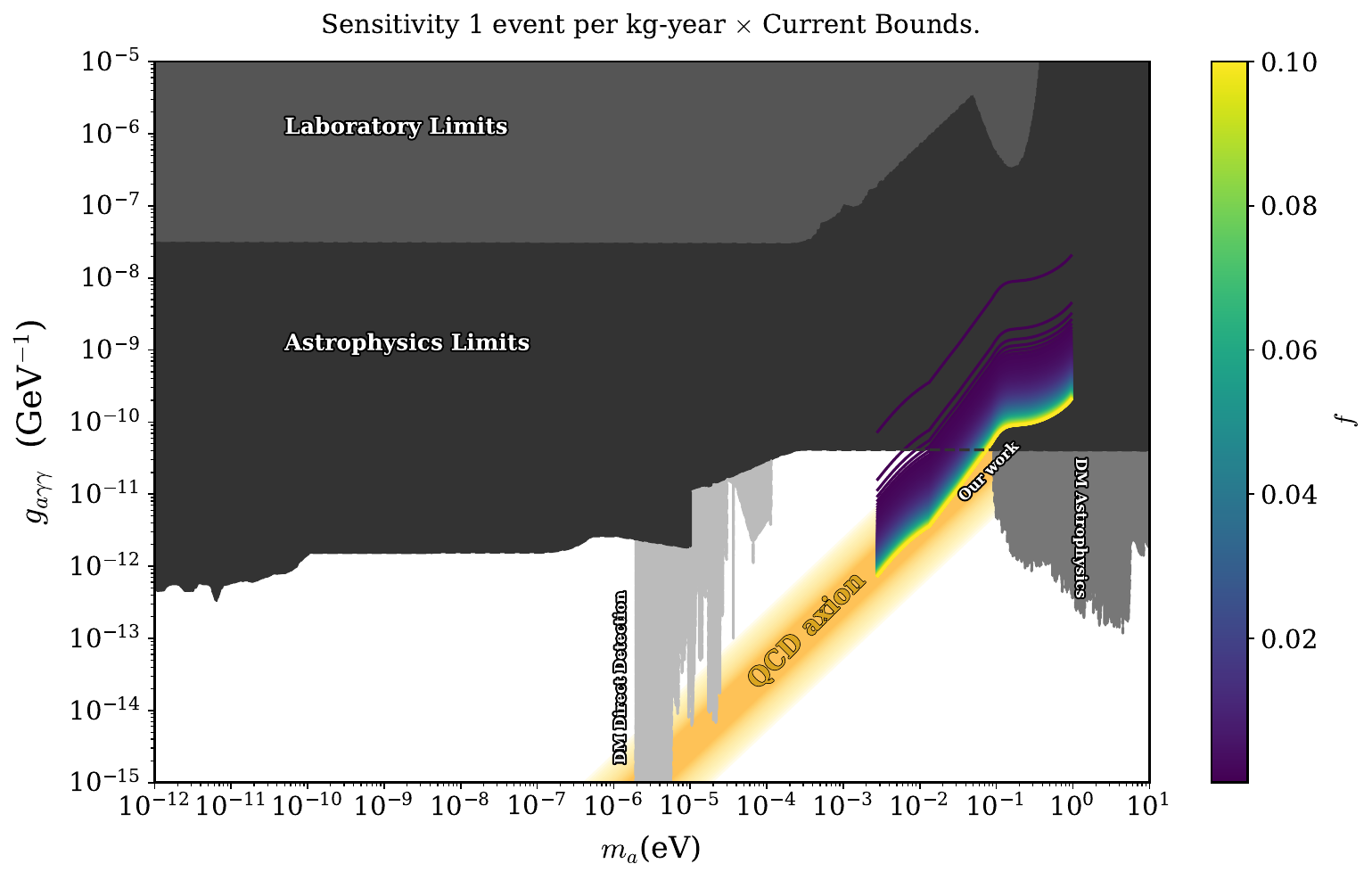}
    \caption{Expected reach to the {\it axion}-photon coupling overlaid with existing bounds in gray. For $m_a < 10^{-1}$ eV our proposal yields a promising sensitivity. We have varied the fraction of boosted dark matter in the range $f=0.01-0.1$.}
    \label{fig:axionkgyearbounds}
\end{figure*}

The reach is found using the rate of interaction of a DM particle in a detector \cite{Cirelli:2024ssz},
\begin{equation}
    R = \frac{\rho_{DM}}{m_{DM}}\frac{1}{\rho}\langle n \sigma v \rangle,
    \label{rate-dm}
\end{equation}
where $\rho_{DM}$ and $m_{DM}$ are the density and the mass of DM respectively, $\rho$ and $n$ are the density and number density of the crystal and $v$ is the DM velocity, which can be regarded as $v \approx 10^{-3}$, however when we substitute Eq. \eqref{crossectionrelations} in Eq. \eqref{rate-dm} there will be no velocity dependence, therefore when working in the framework of BDM to account for the triggering the reaction as explained in Sec. \eqref{tuning}, we only need to add on Eq. \eqref{rate-dm} a new term $f$ that account for the fraction of DM in our Solar System that is boosted and reaches the detector, Eq. \eqref{rate-dm} becomes,
\begin{equation}
    R \simeq f \frac{\rho_{DM}}{m_{DM}}\frac{1}{\rho}\langle n \sigma_{\text{abs}}^\gamma  \rangle,
    \label{rate-fraction}
\end{equation}where $f$ represents the fraction of BDM. We derive the rate by inserting Eq. \eqref{crossectionrelations} in Eq. \eqref{rate-fraction} using absorption data plotted in FIGs. \ref{magneto-chiralimage} and \ref{absorptionDY2}. The expected sensitivity is then calculated for $\omega = m_{A'}$ for a given DM mass in FIG.\ref{expected-sensitivity} for $f=10^{-5} - 10^{-1}$.

The expected sensitivity in FIG.\ref{fig:dykgyearbounds} declines as the amount of BDM available for absorption in the detector decreases . The cusp in the mass region between $10^{-1} - 10^0$ is related to the data in FIG.\ref{magneto-chiralimage}. As can be seen, the absorption gets close enough to zero at $12000 \text{cm}^{-1}$, and given the kinetic mixing dependence on the absorption coefficient scales as $\propto 1/\alpha^{\frac{1}{2}}$, it is to be expected to behave as presented in the FIG.\ref{fig:dykgyearbounds}.  FIG. \ref{fig:dykgyearbounds} explicitly shows the expected sensitivity for an exposure of 1 event per kg-year. We put our findings into perspective with current bounds on {\it dark photons} \cite{Caputo_2021, AxionLimits}. In the mass region of interest, the limits neither come directly from direct detection experiments nor from {\it dark photon} dark matter searches. Nevertheless, the scientific reach of our proposal surpasses that of existing experiments, being able to cover a large region of untouched parameter space.

In particular, the shaded regions are labeled accordingly to the type of experiments in question; $\gamma \to X$ are laboratory limits in $\varepsilon$ coming from light shining through a wall \cite{Parker:2013fxa, Povey:2010hs, Inada:2013tx}, Cavendish-Coulomb \cite{Kroff:2020zhp}; {\it dark photon} as Dark Matter are the limits for the DP corresponding to the totality of DM \cite{Arias:2012az, McDermott:2019lch, Witte:2020rvb, Caputo:2020rnx}, and Stellar Constraints come from astrophysics limits \cite{Dubovsky:2015cca, Wadekar:2019mpc, Bhoonah:2019eyo}; the other colored regions without labels refer to Direct Detection experiments \cite{Schumann:2019eaa, XENON:2023cxc, XENON:2024znc, SuperCDMS:2019jxx} in the mass region of $10^0 - 10^2$, and other laboratory experiments, such as the measurement of Dark-E fields \cite{Godfrey:2021tvs} or radio frequency cavity searches \cite{Nguyen:2019xuh} for the lower mass region: $10^{-8} - 10^{-4}$. The blue-to-yellow region corresponds to our calculations for the two new proposed molecules: Dy$2$ \cite{magneto-chiral} and Dy${}^{\text{III}}$ \cite{near-infrared}. 

As aforementioned, we can reinterpret these results for {\it axions}. Using the absorption spectra for dinuclear dysprosium and assuming
1 event per kg-year of exposure, we ended up probing a region of parameter space already excluded by {\it axion} searches ({\it see Appendix}). That said, we adopted the Mn12-acetate, a complex compound of manganese, and repeated the entire procedure described above and derived the expected sensitivity to {\it axions} as displayed in FIG.\ref{fig:axionkgyearbounds} adopting a magnetic field $B_0 = 1$~T. The shape of the curve is very similar to that of the {\it dark photon} model. The break exhibited in the curve of FIG.\ref{fig:axionkgyearbounds} has to do with the linear dependence in the mass that appears for the coupling, which is not present in the case of the {\it dark photon} model. The projected sensitivity can be improved using a stronger magnetic field.

We have superimposed the existing limits in the literature \cite{AxionLimits}. These bounds stem from laboratory searches \cite{OSQAR:2015qdv, Ehret:2010mh, Betz:2013dza, DellaValle:2015xxa}, astrophysics searches for {\it axions} not as DM \cite{Ayala:2014pea, Dolan:2022kul, CAST:2007jps, Fermi-LAT:2016nkz, Meyer:2016wrm, Meyer:2020vzy, Davies:2022wvj, Noordhuis:2022ljw, Dessert:2021bkv, Dessert:2022yqq, Benabou:2025jcv, Ning:2024eky}, astrophysics bounds for {\it axions} as DM \cite{Janish:2023kvi, Pinetti:2025owq, Saha:2025any, Todarello:2023hdk, Wang:2023imi, Nakayama:2022jza, Carenza:2023qxh, Todarello:2024qci}, and DM direct detection bounds \cite{ADMX:2025vom, Wuensch:1989sa, Bae:2024kmy}.

The purple-to-yellow colored region is a result of our estimate based on Eq. \eqref{eq:axion-dpdm}. The expected reach becomes promising masses $m_a < 10^{-1}$ eV. We emphasize that we can increase the applied magnetic field to improve the sensitivity. From FIG.\ref{fig:axionkgyearbounds}, it is clear that our proposal offers the possibility to probe a sizable region of the QCD {\it axion} parameter space, with leading sensitivity for {\it axion} masses between $2\times 10^{-3}$ and $8\times 10^{-2}$~eV. It is worth noticing that we can push down these projections if we either increase the applied magnetic field or consider a higher fraction of BDM than the one considered in the plots.

In summary, by integrating concepts from particle physics, condensed‑matter physics, and chemistry, we demonstrate that interdisciplinary approaches can enable genuinely new detection strategies and open access to previously unexplored regions of parameter space.

\section{Conclusion \label{Conclusions}} 

The community has been dedicated to testing the compelling WIMP paradigm via accelerators, direct and indirect detection experiments, but null results have been reported thus far. This has motivated the community to explore other dark matter candidates and novel detection mechanisms. In this spirit, we propose to use Single Molecule Magnets as dark matter laboratories, particularly focused on the {\it dark photon} and {\it axion} models.

Single Molecule Magnets function as frustrated systems stabilized in metastable states for relatively long periods. Dark-matter absorption triggers accelerated relaxation from these states, producing detectable magnetic avalanches. Using dysprosium- and manganese-based molecules, we computed reaction rates for dark-photon and axion interactions, deriving projected sensitivities. Our analysis demonstrates that dysprosium SMMs improve sensitivity to dark photons by orders of magnitude, while manganese systems probe previously inaccessible regions of the QCD axion parameter space.

Our work builds a bridge between chemistry, condensed-matter physics, and particle physics for dark-matter detection. Future characterization of magnetic materials across infrared to ultraviolet frequencies will critically enhance the sensitivity of this approach and pave the road for the potential detection of dark matter particles.

\section{Acknowledgments}

The authors thank Raimundo Silva and Jacinto P. Neto for insightful discussions. FSQ thanks CERN and the Max Planck Institute for Nuclear Physics for the hospitality where this work was partly developed. JA is funded by the {\it Coordenação de Aperfeiçoamento de Pessoal de Nível Superior (CAPES)} and CNPq grants 88887.989987/2024-00 and 403521/2024-6. FSQ acknowledges support from Simons Foundation (Award Number:1023171-RC), CNPq 403521/2024-6, 408295/2021-0, 403521/2024-6, 406919/2025-9, 351851/2025-9, the FAPESP Grants 2021/01089-1, 2023/01197-4, ICTP-SAIFR 2021/14335-0, and the ANID-Millennium Science Initiative Program ICN2019\_044. This work is partially funded by FINEP under project 213/2024 and was carried out in part through the IIP cluster {\it bulletcluster}

\section{Appendix A}

In this section, we present the sensitivity with an exposure of 1 event per kg-day to both {\it dark photon} and {\it axion} dark matter particles. These plots are an extension of the results are exhibited in FIGs.\ref{fig:dykgyearbounds}-\ref{fig:axionkgyearbounds} but offer a clearer vision of the scientific reach in comparison with existing bounds and the role of each compound. 

\begin{figure}[h]
    \centering
    \begin{subfigure}[b]{0.48\textwidth}
        \centering
        \includegraphics[width=\textwidth]{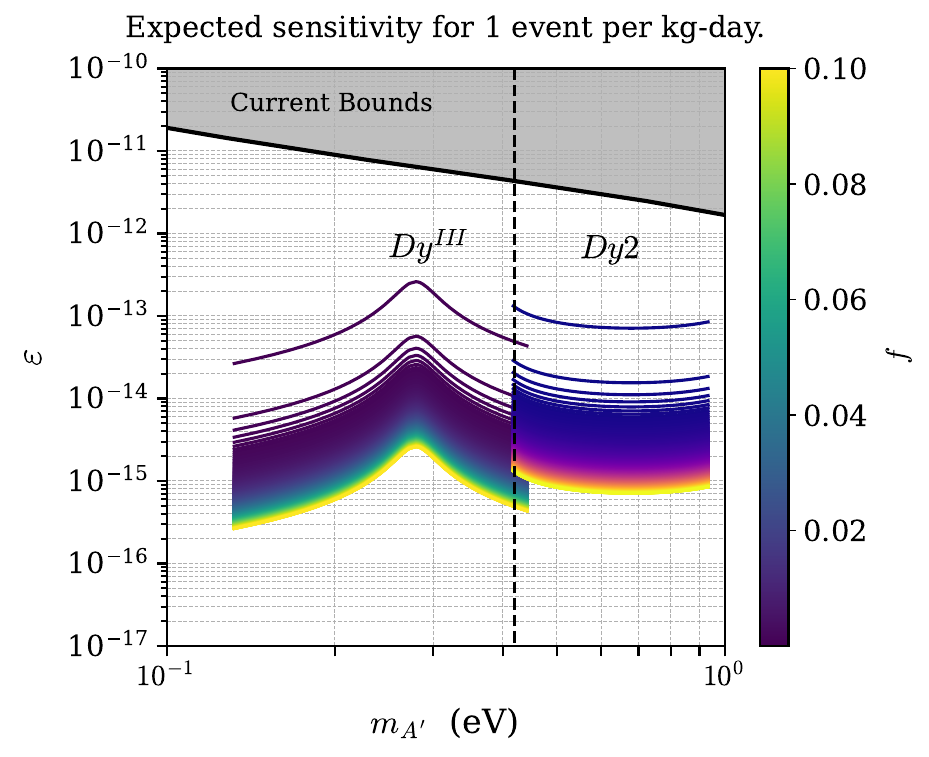}
        \caption{Sensitivity for the kinetic mixing with an exposure of 1 event per kg-day.}
        \label{expectedsensitivitykgday}
    \end{subfigure}
    \hfill
    \begin{subfigure}[b]{0.48\textwidth}
        \centering
        \includegraphics[width=\textwidth]{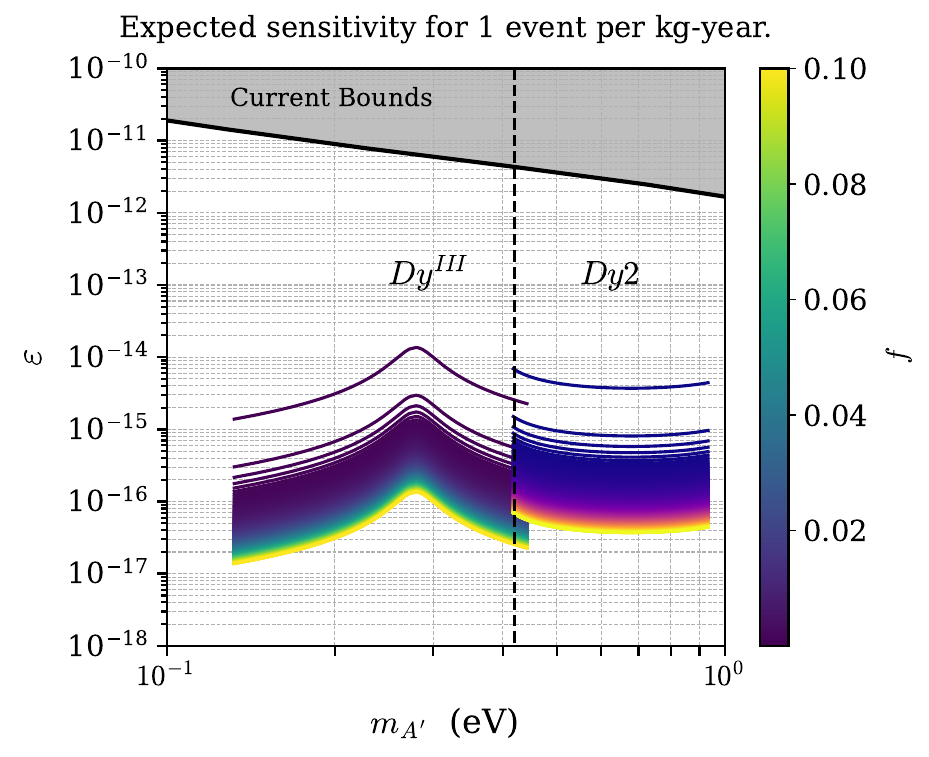}
        \caption{Sensitivity for the kinetic mixing with one event per kg-year exposure.}
        \label{expected-sensitivity-kgyear}
    \end{subfigure}
    \caption{Expected sensitivities for different exposure times for the {\it dark photon} model assuming: (a) 1 event per kg-day; (b) one event per kg-year.}
    \label{expected-sensitivity}
\end{figure}

In FIGs.\ref{expectedsensitivitykgday}-\ref{expected-sensitivity-kgyear} we display the sensitivity for two different exposures using dysprosium molecules. Notoriously, with a larger exposure, a greater reach is obtained.  This behavior is expected since the kinetic mixing scales as $(\text{kg-time})^{-1/2}$. These findings were derived using Eq. \eqref{rate-fraction} for $f=0.01- 0.1$. From FIGs.\ref{expectedsensitivitykgday}-\ref{expected-sensitivity-kgyear}, we solidly show that such molecules offer a powerful platform for probing physics beyond the Standard Model. 

\begin{figure}[h]
    \centering
    \includegraphics[width=\linewidth]{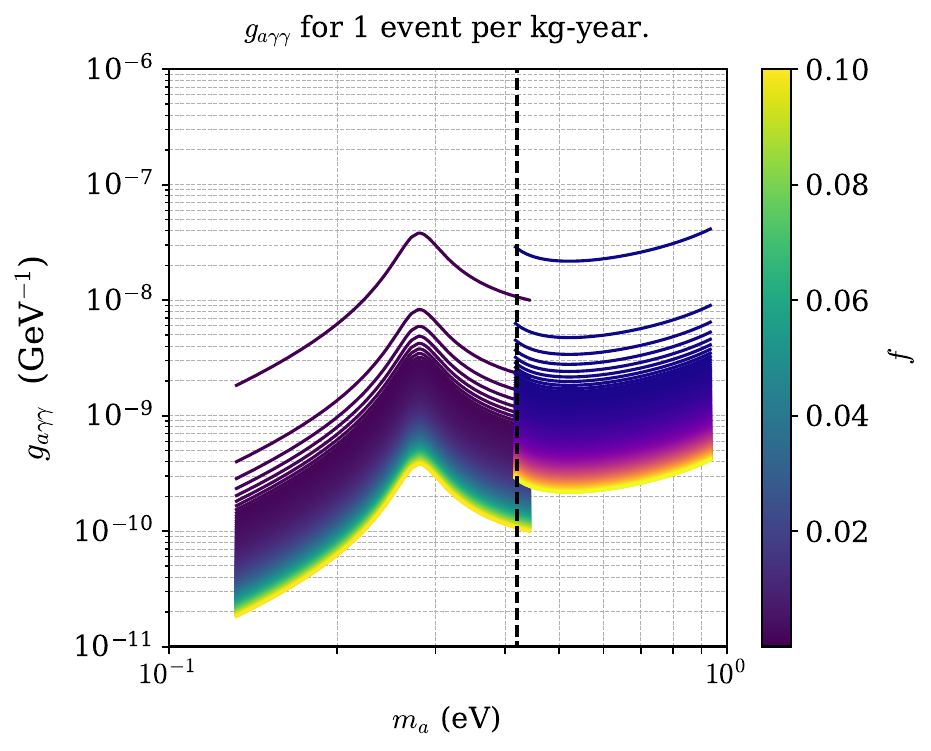}
    \caption{Expected reach to {\it axion}-photon coupling for the dysprosium molecules assuming 1 event per kg-year.}
    \label{fig:axion-photon-dy}
\end{figure}

\begin{figure}[h!]
    \centering
    \begin{subfigure}[b]{0.48\textwidth}
        \centering
        \includegraphics[width=\textwidth]{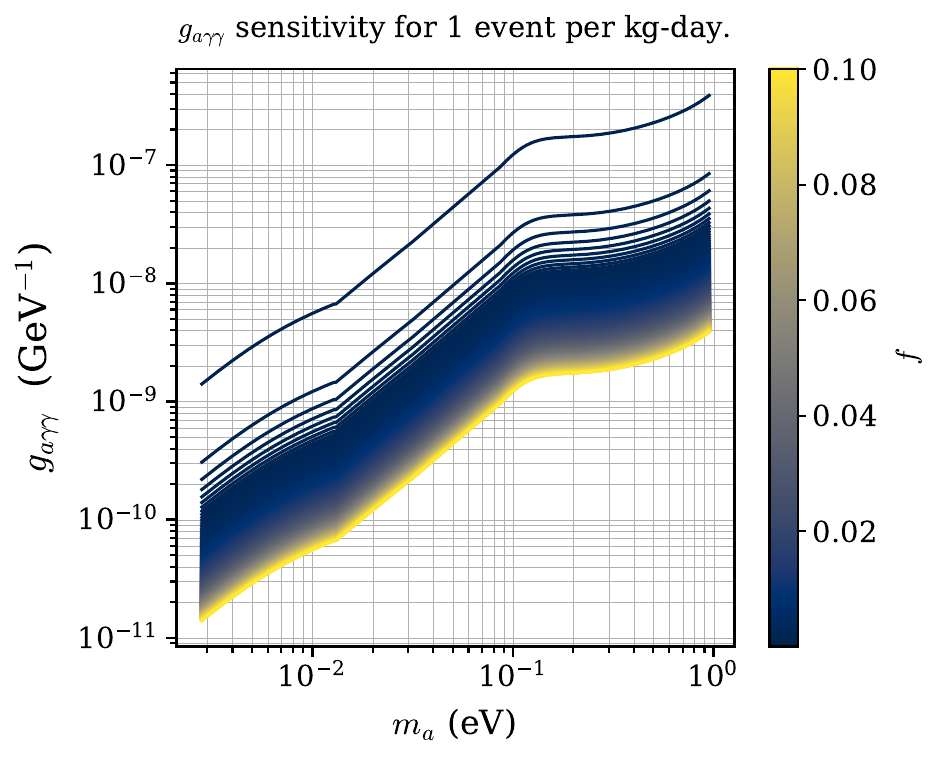}
        \caption{Sensitivity for the {\it axion}-photon coupling with an exposure of 1 event per kg-day.}
        \label{fig:axionkgday}
    \end{subfigure}
    \hfill
    \begin{subfigure}[b]{0.48\textwidth}
        \centering
        \includegraphics[width=\textwidth]{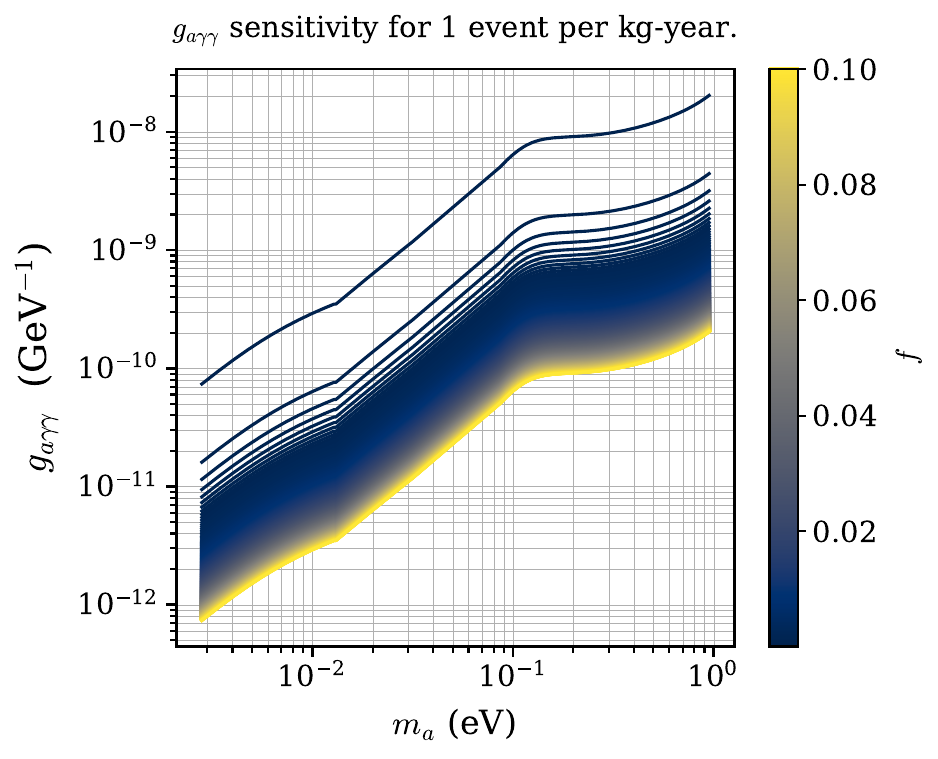}
        \caption{Sensitivity for the {\it axion}-photon coupling with one event per kg-year exposure.}
        \label{fig:axionkgyear}
    \end{subfigure}
    \caption{Projected reach of a manganese compound to the {\it axion}-photon coupling using 1 one event per kg-day (figure a), and 1 one event per kg-day (figure b).}
    \label{fig:axion-photon-sensitivity}
\end{figure}

Using the same molecules, we also derived the sensitivity to the {\it axion}-photon coupling. The results are summarized in FIG.\ref{fig:axion-photon-dy} for an exposure of 1 event per kg-year. The parameter space covered in FIG.\ref{fig:axion-photon-dy} has already been excluded. Thus, we decided to use a different molecule, Mn12-acetate, which is a complex compound of manganese, and repeat the same procedure. The expected reach is shown in FIGs.\ref{fig:axionkgday}-\ref{fig:axionkgyear} for different exposures, similar to the previous figures. Our finding is based on Eq. \eqref{eq:axion-dpdm} and the sensitivity projected for kinetic mixing using  Mn12-acetate \cite{Bunting_2017}. The shape of the curve is very similar to that of the {\it dark photon} model. The small tilt in FIGs.\ref{fig:axionkgday}-\ref{fig:axionkgyear} is due to the linear dependence in the mass that appears for the coupling, which is not present in the case of the kinetic mixing. We have adopted a magnetic field  $B_0 = 1$T, but a better sensitivity can be achieved with larger magnetic fields.  

Our findings were based on two compounds, and for this reason, there is a visible break in the curves. In particular, we observe that dy2 covers larger masses compared to the $Dy^{III}$. 

\bibliography{references}

@article{Wang:2025ztb,
    author = "Wang, Jin-Wei",
    title = "{Blazar-boosted dark matter: Novel signatures via elastic and inelastic scattering}",
    eprint = "2503.22105",
    archivePrefix = "arXiv",
    primaryClass = "hep-ph",
    reportNumber = "Phys. Rev. D 112, 055004",
    doi = "10.1103/wr32-3g38",
    journal = "Phys. Rev. D",
    volume = "112",
    number = "5",
    pages = "055004",
    year = "2025"
}

@article{Cirelli:2024ssz,
    author = "Cirelli, Marco and Strumia, Alessandro and Zupan, Jure",
    title = "{Dark Matter}",
    eprint = "2406.01705",
    archivePrefix = "arXiv",
    primaryClass = "hep-ph",
    month = "6",
    year = "2024"
}

@article{mn12data2,
  title = {Diffuse optical excitations in ${\mathrm{Mn}}_{12}$-acetate},
  author = {Oppenheimer, S. M. and Sushkov, A. B. and Musfeldt, J. L. and Achey, R. M. and Dalal, N. S.},
  journal = {Phys. Rev. B},
  volume = {65},
  issue = {5},
  pages = {054419},
  numpages = {5},
  year = {2002},
  month = {Jan},
  publisher = {American Physical Society},
  doi = {10.1103/PhysRevB.65.054419},
  url = {https://link.aps.org/doi/10.1103/PhysRevB.65.054419}
}

@article{Schumann:2019eaa,
    author = "Schumann, Marc",
    title = "{Direct Detection of WIMP Dark Matter: Concepts and Status}",
    eprint = "1903.03026",
    archivePrefix = "arXiv",
    primaryClass = "astro-ph.CO",
    doi = "10.1088/1361-6471/ab2ea5",
    journal = "J. Phys. G",
    volume = "46",
    number = "10",
    pages = "103003",
    year = "2019"
}

@article{Green:2024bam,
    author = "Green, Anne M.",
    title = "{Primordial black holes as a dark matter candidate - a brief overview}",
    eprint = "2402.15211",
    archivePrefix = "arXiv",
    primaryClass = "astro-ph.CO",
    doi = "10.1016/j.nuclphysb.2024.116494",
    journal = "Nucl. Phys. B",
    volume = "1003",
    pages = "116494",
    year = "2024"
}

@article{mn12data1,
  title = {Magnetic field effects on the far-infrared absorption in ${\mathrm{Mn}}_{12}$-acetate},
  author = {Sushkov, A. B. and Jones, B. R. and Musfeldt, J. L. and Wang, Y. J. and Achey, R. M. and Dalal, N. S.},
  journal = {Phys. Rev. B},
  volume = {63},
  issue = {21},
  pages = {214408},
  numpages = {6},
  year = {2001},
  month = {May},
  publisher = {American Physical Society},
  doi = {10.1103/PhysRevB.63.214408},
  url = {https://link.aps.org/doi/10.1103/PhysRevB.63.214408}
}

@article{Essig:2017kqs,
    author = "Essig, Rouven and Volansky, Tomer and Yu, Tien-Tien",
    title = "{New Constraints and Prospects for sub-GeV Dark Matter Scattering off Electrons in Xenon}",
    eprint = "1703.00910",
    archivePrefix = "arXiv",
    primaryClass = "hep-ph",
    reportNumber = "CERN-TH-2017-042, YITP-SB-17-09",
    doi = "10.1103/PhysRevD.96.043017",
    journal = "Phys. Rev. D",
    volume = "96",
    number = "4",
    pages = "043017",
    year = "2017"
}

@article{Essig:2012yx,
    author = "Essig, Rouven and Manalaysay, Aaron and Mardon, Jeremy and Sorensen, Peter and Volansky, Tomer",
    title = "{First Direct Detection Limits on sub-GeV Dark Matter from XENON10}",
    eprint = "1206.2644",
    archivePrefix = "arXiv",
    primaryClass = "astro-ph.CO",
    reportNumber = "YITP-SB-01-12",
    doi = "10.1103/PhysRevLett.109.021301",
    journal = "Phys. Rev. Lett.",
    volume = "109",
    pages = "021301",
    year = "2012"
}

@article{Herrera:2016hlp,
    author = "Herrera, Felipe and Spano, Frank C.",
    title = "{Dark Vibronic Polaritons and the Spectroscopy of Organic Microcavities}",
    eprint = "1610.04252",
    archivePrefix = "arXiv",
    primaryClass = "quant-ph",
    doi = "10.1103/PhysRevLett.118.223601",
    journal = "Phys. Rev. Lett.",
    volume = "118",
    pages = "223601",
    year = "2017"
}

@article{Agashe:2014yua,
    author = "Agashe, Kaustubh and Cui, Yanou and Necib, Lina and Thaler, Jesse",
    title = "{(In)direct Detection of Boosted Dark Matter}",
    eprint = "1405.7370",
    archivePrefix = "arXiv",
    primaryClass = "hep-ph",
    reportNumber = "MIT-CTP-4538, UMD-PP-014-005",
    doi = "10.1088/1475-7516/2014/10/062",
    journal = "JCAP",
    volume = "10",
    pages = "062",
    year = "2014"
}

@article{Ghosh:2024dqw,
    author = "Ghosh, Dilip Kumar and Gupta, Tushar and Heikinheimo, Matti and Huitu, Katri and Jeesun, Sk",
    title = "{Boosted dark matter driven by cosmic rays and diffuse supernova neutrinos}",
    eprint = "2411.11973",
    archivePrefix = "arXiv",
    primaryClass = "hep-ph",
    doi = "10.1103/PhysRevD.111.063019",
    journal = "Phys. Rev. D",
    volume = "111",
    number = "6",
    pages = "063019",
    year = "2025"
}

@book{Fabbrichesi_2021,
   title={The Physics of the Dark Photon: A Primer},
   ISBN={9783030625191},
   ISSN={2191-5431},
   url={http://dx.doi.org/10.1007/978-3-030-62519-1},
   DOI={10.1007/978-3-030-62519-1},
   journal={SpringerBriefs in Physics},
   publisher={Springer International Publishing},
   author={Fabbrichesi, Marco and Gabrielli, Emidio and Lanfranchi, Gaia},
   year={2021} }

@article{Bunting_2017,
   title={Magnetic bubble chambers and sub-GeV dark matter direct detection},
   volume={95},
   ISSN={2470-0029},
   url={http://dx.doi.org/10.1103/PhysRevD.95.095001},
   DOI={10.1103/physrevd.95.095001},
   number={9},
   journal={Physical Review D},
   publisher={American Physical Society (APS)},
   author={Bunting, Philip C. and Gratta, Giorgio and Melia, Tom and Rajendran, Surjeet},
   year={2017},
   month=may }

@article{Godfrey:2021tvs,
    author = "Godfrey, Benjamin and others",
    title = "{Search for dark photon dark matter: Dark E field radio pilot experiment}",
    eprint = "2101.02805",
    archivePrefix = "arXiv",
    primaryClass = "physics.ins-det",
    doi = "10.1103/PhysRevD.104.012013",
    journal = "Phys. Rev. D",
    volume = "104",
    number = "1",
    pages = "012013",
    year = "2021"
}

@article{Povey:2010hs,
    author = "Povey, Rhys and Hartnett, John and Tobar, Michael",
    title = "{Microwave cavity light shining through a wall optimization and experiment}",
    eprint = "1003.0964",
    archivePrefix = "arXiv",
    primaryClass = "hep-ex",
    doi = "10.1103/PhysRevD.82.052003",
    journal = "Phys. Rev. D",
    volume = "82",
    pages = "052003",
    year = "2010"
}

@article{Caputo:2020rnx,
    author = "Caputo, Andrea and Liu, Hongwan and Mishra-Sharma, Siddharth and Ruderman, Joshua T.",
    title = "{Modeling Dark Photon Oscillations in Our Inhomogeneous Universe}",
    eprint = "2004.06733",
    archivePrefix = "arXiv",
    primaryClass = "astro-ph.CO",
    doi = "10.1103/PhysRevD.102.103533",
    journal = "Phys. Rev. D",
    volume = "102",
    number = "10",
    pages = "103533",
    year = "2020"
}

@article{Dubovsky:2015cca,
    author = "Dubovsky, Sergei and Hern{\'a}ndez-Chifflet, Guzm{\'a}n",
    title = "{Heating up the Galaxy with Hidden Photons}",
    eprint = "1509.00039",
    archivePrefix = "arXiv",
    primaryClass = "hep-ph",
    doi = "10.1088/1475-7516/2015/12/054",
    journal = "JCAP",
    volume = "12",
    pages = "054",
    year = "2015"
}

@article{Wadekar:2019mpc,
    author = "Wadekar, Digvijay and Farrar, Glennys R.",
    title = "{Gas-rich dwarf galaxies as a new probe of dark matter interactions with ordinary matter}",
    eprint = "1903.12190",
    archivePrefix = "arXiv",
    primaryClass = "hep-ph",
    doi = "10.1103/PhysRevD.103.123028",
    journal = "Phys. Rev. D",
    volume = "103",
    number = "12",
    pages = "123028",
    year = "2021"
}

@article{OSQAR:2015qdv,
    author = "Ballou, R. and others",
    collaboration = "OSQAR",
    title = "{New exclusion limits on scalar and pseudoscalar axionlike particles from light shining through a wall}",
    eprint = "1506.08082",
    archivePrefix = "arXiv",
    primaryClass = "hep-ex",
    doi = "10.1103/PhysRevD.92.092002",
    journal = "Phys. Rev. D",
    volume = "92",
    number = "9",
    pages = "092002",
    year = "2015"
}

@article{Dolan:2022kul,
    author = "Dolan, Matthew J. and Hiskens, Frederick J. and Volkas, Raymond R.",
    title = "{Advancing globular cluster constraints on the axion-photon coupling}",
    eprint = "2207.03102",
    archivePrefix = "arXiv",
    primaryClass = "hep-ph",
    doi = "10.1088/1475-7516/2022/10/096",
    journal = "JCAP",
    volume = "10",
    pages = "096",
    year = "2022"
}

@article{Ayala:2014pea,
    author = "Ayala, Adrian and Dom{\'\i}nguez, Inma and Giannotti, Maurizio and Mirizzi, Alessandro and Straniero, Oscar",
    title = "{Revisiting the bound on axion-photon coupling from Globular Clusters}",
    eprint = "1406.6053",
    archivePrefix = "arXiv",
    primaryClass = "astro-ph.SR",
    doi = "10.1103/PhysRevLett.113.191302",
    journal = "Phys. Rev. Lett.",
    volume = "113",
    number = "19",
    pages = "191302",
    year = "2014"
}

@article{CAST:2007jps,
    author = "Andriamonje, S. and others",
    collaboration = "CAST",
    title = "{An Improved limit on the axion-photon coupling from the CAST experiment}",
    eprint = "hep-ex/0702006",
    archivePrefix = "arXiv",
    doi = "10.1088/1475-7516/2007/04/010",
    journal = "JCAP",
    volume = "04",
    pages = "010",
    year = "2007"
}

@article{Fermi-LAT:2016nkz,
    author = "Ajello, M. and others",
    collaboration = "Fermi-LAT",
    title = "{Search for Spectral Irregularities due to Photon{\textendash}Axionlike-Particle Oscillations with the Fermi Large Area Telescope}",
    eprint = "1603.06978",
    archivePrefix = "arXiv",
    primaryClass = "astro-ph.HE",
    doi = "10.1103/PhysRevLett.116.161101",
    journal = "Phys. Rev. Lett.",
    volume = "116",
    number = "16",
    pages = "161101",
    year = "2016"
}

@article{Davies:2022wvj,
    author = "Davies, James and Meyer, Manuel and Cotter, Garret",
    title = "{Constraints on axionlike particles from a combined analysis of three flaring Fermi flat-spectrum radio quasars}",
    eprint = "2211.03414",
    archivePrefix = "arXiv",
    primaryClass = "astro-ph.HE",
    doi = "10.1103/PhysRevD.107.083027",
    journal = "Phys. Rev. D",
    volume = "107",
    number = "8",
    pages = "083027",
    year = "2023"
}

@article{Noordhuis:2022ljw,
    author = "Noordhuis, Dion and Prabhu, Anirudh and Witte, Samuel J. and Chen, Alexander Y. and Cruz, F{\'a}bio and Weniger, Christoph",
    title = "{Novel Constraints on Axions Produced in Pulsar Polar-Cap Cascades}",
    eprint = "2209.09917",
    archivePrefix = "arXiv",
    primaryClass = "hep-ph",
    doi = "10.1103/PhysRevLett.131.111004",
    journal = "Phys. Rev. Lett.",
    volume = "131",
    number = "11",
    pages = "111004",
    year = "2023"
}

@article{Dessert:2021bkv,
    author = "Dessert, Christopher and Long, Andrew J. and Safdi, Benjamin R.",
    title = "{No Evidence for Axions from Chandra Observation of the Magnetic White Dwarf RE J0317-853}",
    eprint = "2104.12772",
    archivePrefix = "arXiv",
    primaryClass = "hep-ph",
    doi = "10.1103/PhysRevLett.128.071102",
    journal = "Phys. Rev. Lett.",
    volume = "128",
    number = "7",
    pages = "071102",
    year = "2022"
}

@article{Dessert:2022yqq,
    author = "Dessert, Christopher and Dunsky, David and Safdi, Benjamin R.",
    title = "{Upper limit on the axion-photon coupling from magnetic white dwarf polarization}",
    eprint = "2203.04319",
    archivePrefix = "arXiv",
    primaryClass = "hep-ph",
    doi = "10.1103/PhysRevD.105.103034",
    journal = "Phys. Rev. D",
    volume = "105",
    number = "10",
    pages = "103034",
    year = "2022"
}

@article{Janish:2023kvi,
    author = "Janish, Ryan and Pinetti, Elena",
    title = "{Hunting Dark Matter Lines in the Infrared Background with the James Webb Space Telescope}",
    eprint = "2310.15395",
    archivePrefix = "arXiv",
    primaryClass = "hep-ph",
    reportNumber = "FERMILAB-PUB-23-633-T",
    doi = "10.1103/PhysRevLett.134.071002",
    journal = "Phys. Rev. Lett.",
    volume = "134",
    number = "7",
    pages = "071002",
    year = "2025"
}

@article{Pinetti:2025owq,
    author = "Pinetti, Elena",
    title = "{First constraints on QCD axion dark matter using James Webb Space Telescope observations}",
    eprint = "2503.11753",
    archivePrefix = "arXiv",
    primaryClass = "hep-ph",
    reportNumber = "FERMILAB-PUB-25-0166-V",
    month = "3",
    year = "2025"
}

@article{Saha:2025any,
    author = "Saha, Akash Kumar and Bouri, Subhadip and Das, Anirban and Dubey, Abhishek and Laha, Ranjan",
    title = "{Shedding Infrared Light on QCD Axion and ALP Dark Matter with JWST}",
    eprint = "2503.14582",
    archivePrefix = "arXiv",
    primaryClass = "hep-ph",
    month = "3",
    year = "2025"
}

@article{Todarello:2023hdk,
    author = "Todarello, Elisa and Regis, Marco and Reynoso-Cordova, Javier and Taoso, Marco and Vaz, Daniel and Brinchmann, Jarle and Steinmetz, Matthias and Zoutendijke, Sebastiaan L.",
    title = "{Robust bounds on ALP dark matter from dwarf spheroidal galaxies in the optical MUSE-Faint survey}",
    eprint = "2307.07403",
    archivePrefix = "arXiv",
    primaryClass = "astro-ph.CO",
    doi = "10.1088/1475-7516/2024/05/043",
    journal = "JCAP",
    volume = "05",
    pages = "043",
    year = "2024"
}

@article{Wang:2023imi,
    author = "Wang, Hanyue and others",
    title = "{Spectroscopic search for optical emission lines from dark matter decay}",
    eprint = "2311.05476",
    archivePrefix = "arXiv",
    primaryClass = "astro-ph.CO",
    doi = "10.1103/PhysRevD.110.103007",
    journal = "Phys. Rev. D",
    volume = "110",
    number = "10",
    pages = "103007",
    year = "2024"
}

@article{Nakayama:2022jza,
    author = "Nakayama, Kazunori and Yin, Wen",
    title = "{Anisotropic cosmic optical background bound for decaying dark matter in light of the LORRI anomaly}",
    eprint = "2205.01079",
    archivePrefix = "arXiv",
    primaryClass = "hep-ph",
    reportNumber = "TU-1154",
    doi = "10.1103/PhysRevD.106.103505",
    journal = "Phys. Rev. D",
    volume = "106",
    number = "10",
    pages = "103505",
    year = "2022"
}

@article{Carenza:2023qxh,
    author = "Carenza, Pierluca and Lucente, Giuseppe and Vitagliano, Edoardo",
    title = "{Probing the blue axion with cosmic optical background anisotropies}",
    eprint = "2301.06560",
    archivePrefix = "arXiv",
    primaryClass = "hep-ph",
    doi = "10.1103/PhysRevD.107.083032",
    journal = "Phys. Rev. D",
    volume = "107",
    number = "8",
    pages = "083032",
    year = "2023"
}

@article{Todarello:2024qci,
    author = "Todarello, Elisa and Regis, Marco",
    title = "{Bounds on axions-like particles shining in the ultra-violet}",
    eprint = "2412.02543",
    archivePrefix = "arXiv",
    primaryClass = "hep-ph",
    doi = "10.1088/1475-7516/2025/05/070",
    journal = "JCAP",
    volume = "05",
    pages = "070",
    year = "2025"
}

@article{ADMX:2025vom,
    author = "Carosi, G. and others",
    collaboration = "ADMX",
    title = "{Search for Axion Dark Matter from 1.1 to 1.3~GHz with ADMX}",
    eprint = "2504.07279",
    archivePrefix = "arXiv",
    primaryClass = "hep-ex",
    reportNumber = "FERMILAB-PUB-25-0298-PPD",
    doi = "10.1103/d7mg-6sqq",
    journal = "Phys. Rev. Lett.",
    volume = "135",
    number = "19",
    pages = "191001",
    year = "2025"
}

@article{Wuensch:1989sa,
    author = "Wuensch, Walter and De Panfilis-Wuensch, S. and Semertzidis, Y. K. and Rogers, J. T. and Melissinos, A. C. and Halama, H. J. and Moskowitz, B. E. and Prodell, A. G. and Fowler, W. B. and Nezrick, F. A.",
    title = "{Results of a Laboratory Search for Cosmic Axions and Other Weakly Coupled Light Particles}",
    reportNumber = "FERMILAB-PUB-89-185-E, BNL-43010",
    doi = "10.1103/PhysRevD.40.3153",
    journal = "Phys. Rev. D",
    volume = "40",
    pages = "3153",
    year = "1989"
}

@article{Bae:2024kmy,
    author = "Bae, Sungjae and Jeong, Junu and Kim, Younggeun and Youn, SungWoo and Park, Heejun and Seong, Taehyeon and Oh, Seongjeong and Semertzidis, Yannis K.",
    title = "{Search for Dark Matter Axions with Tunable TM020 Mode}",
    eprint = "2403.13390",
    archivePrefix = "arXiv",
    primaryClass = "hep-ex",
    doi = "10.1103/PhysRevLett.133.211803",
    journal = "Phys. Rev. Lett.",
    volume = "133",
    number = "21",
    pages = "211803",
    year = "2024"
}

@article{Benabou:2025jcv,
    author = "Benabou, Joshua N. and Dessert, Christopher and Patra, Kishore C. and Brink, Thomas G. and Zheng, WeiKang and Filippenko, Alexei V. and Safdi, Benjamin R.",
    title = "{Search for Axions in Magnetic White Dwarf Polarization at Lick and Keck Observatories}",
    eprint = "2504.12377",
    archivePrefix = "arXiv",
    primaryClass = "hep-ph",
    month = "4",
    year = "2025"
}

@article{Ning:2024eky,
    author = "Ning, Orion and Safdi, Benjamin R.",
    title = "{Leading Axion-Photon Sensitivity with NuSTAR Observations of M82 and M87}",
    eprint = "2404.14476",
    archivePrefix = "arXiv",
    primaryClass = "hep-ph",
    doi = "10.1103/PhysRevLett.134.171003",
    journal = "Phys. Rev. Lett.",
    volume = "134",
    number = "17",
    pages = "171003",
    year = "2025"
}

@article{Meyer:2020vzy,
    author = "Meyer, Manuel and Petrushevska, Tanja",
    title = "{Search for Axionlike-Particle-Induced Prompt $\gamma$-Ray Emission from Extragalactic Core-Collapse Supernovae with the $Fermi$ Large Area Telescope}",
    eprint = "2006.06722",
    archivePrefix = "arXiv",
    primaryClass = "astro-ph.HE",
    doi = "10.1103/PhysRevLett.124.231101",
    journal = "Phys. Rev. Lett.",
    volume = "124",
    number = "23",
    pages = "231101",
    year = "2020",
    note = "[Erratum: Phys.Rev.Lett. 125, 119901 (2020)]"
}

@article{Meyer:2016wrm,
    author = "Meyer, M. and Giannotti, M. and Mirizzi, A. and Conrad, J. and S{\'a}nchez-Conde, M. A.",
    title = "{Fermi Large Area Telescope as a Galactic Supernovae Axionscope}",
    eprint = "1609.02350",
    archivePrefix = "arXiv",
    primaryClass = "astro-ph.HE",
    doi = "10.1103/PhysRevLett.118.011103",
    journal = "Phys. Rev. Lett.",
    volume = "118",
    number = "1",
    pages = "011103",
    year = "2017"
}

@article{DellaValle:2015xxa,
    author = "Della Valle, Federico and Ejlli, Aldo and Gastaldi, Ugo and Messineo, Giuseppe and Milotti, Edoardo and Pengo, Ruggero and Ruoso, Giuseppe and Zavattini, Guido",
    title = "{The PVLAS experiment: measuring vacuum magnetic birefringence and dichroism with a birefringent Fabry{\textendash}Perot cavity}",
    eprint = "1510.08052",
    archivePrefix = "arXiv",
    primaryClass = "physics.optics",
    doi = "10.1140/epjc/s10052-015-3869-8",
    journal = "Eur. Phys. J. C",
    volume = "76",
    number = "1",
    pages = "24",
    year = "2016"
}

@article{Betz:2013dza,
    author = "Betz, M. and Caspers, F. and Gasior, M. and Thumm, M. and Rieger, S. W.",
    title = "{First results of the CERN Resonant Weakly Interacting sub-eV Particle Search (CROWS)}",
    eprint = "1310.8098",
    archivePrefix = "arXiv",
    primaryClass = "physics.ins-det",
    doi = "10.1103/PhysRevD.88.075014",
    journal = "Phys. Rev. D",
    volume = "88",
    number = "7",
    pages = "075014",
    year = "2013"
}

@article{Ehret:2010mh,
    author = "Ehret, Klaus and others",
    title = "{New ALPS Results on Hidden-Sector Lightweights}",
    eprint = "1004.1313",
    archivePrefix = "arXiv",
    primaryClass = "hep-ex",
    reportNumber = "DESY-10-030, MPP-2010-27",
    doi = "10.1016/j.physletb.2010.04.066",
    journal = "Phys. Lett. B",
    volume = "689",
    pages = "149--155",
    year = "2010"
}

@article{Bhoonah:2019eyo,
    author = "Bhoonah, Amit and Bramante, Joseph and Song, Ningqiang",
    title = "{Superradiant Searches for Dark Photons in Two Stage Atomic Transitions}",
    eprint = "1909.07387",
    archivePrefix = "arXiv",
    primaryClass = "hep-ph",
    doi = "10.1103/PhysRevD.101.055040",
    journal = "Phys. Rev. D",
    volume = "101",
    number = "5",
    pages = "055040",
    year = "2020"
}

@article{Witte:2020rvb,
    author = "Witte, Samuel J. and Rosauro-Alcaraz, Salvador and McDermott, Samuel D. and Poulin, Vivian",
    title = "{Dark photon dark matter in the presence of inhomogeneous structure}",
    eprint = "2003.13698",
    archivePrefix = "arXiv",
    primaryClass = "astro-ph.CO",
    reportNumber = "FERMILAB-PUB-20-121-T",
    doi = "10.1007/JHEP06(2020)132",
    journal = "JHEP",
    volume = "06",
    pages = "132",
    year = "2020"
}

@article{McDermott:2019lch,
    author = "McDermott, Samuel D. and Witte, Samuel J.",
    title = "{Cosmological evolution of light dark photon dark matter}",
    eprint = "1911.05086",
    archivePrefix = "arXiv",
    primaryClass = "hep-ph",
    reportNumber = "FERMILAB-PUB-19-565-T",
    doi = "10.1103/PhysRevD.101.063030",
    journal = "Phys. Rev. D",
    volume = "101",
    number = "6",
    pages = "063030",
    year = "2020"
}

@article{Arias:2012az,
    author = "Arias, Paola and Cadamuro, Davide and Goodsell, Mark and Jaeckel, Joerg and Redondo, Javier and Ringwald, Andreas",
    title = "{WISPy Cold Dark Matter}",
    eprint = "1201.5902",
    archivePrefix = "arXiv",
    primaryClass = "hep-ph",
    reportNumber = "DESY-11-226, MPP-2011-140, CERN-PH-TH-2011-323, IPPP-11-80, DCPT-11-160",
    doi = "10.1088/1475-7516/2012/06/013",
    journal = "JCAP",
    volume = "06",
    pages = "013",
    year = "2012"
}

@article{SuperCDMS:2019jxx,
    author = "Aralis, T. and others",
    collaboration = "SuperCDMS",
    title = "{Constraints on dark photons and axionlike particles from the SuperCDMS Soudan experiment}",
    eprint = "1911.11905",
    archivePrefix = "arXiv",
    primaryClass = "hep-ex",
    reportNumber = "FERMILAB-PUB-19-666-AE-TD",
    doi = "10.1103/PhysRevD.101.052008",
    journal = "Phys. Rev. D",
    volume = "101",
    number = "5",
    pages = "052008",
    year = "2020",
    note = "[Erratum: Phys.Rev.D 103, 039901 (2021)]"
}

@article{Kroff:2020zhp,
    author = "Kroff, D. and Malta, P. C.",
    title = "{Constraining hidden photons via atomic force microscope measurements and the Plimpton-Lawton experiment}",
    eprint = "2008.02209",
    archivePrefix = "arXiv",
    primaryClass = "hep-ph",
    doi = "10.1103/PhysRevD.102.095015",
    journal = "Phys. Rev. D",
    volume = "102",
    number = "9",
    pages = "095015",
    year = "2020"
}

@article{Inada:2013tx,
    author = "Inada, T. and Namba, T. and Asai, S. and Kobayashi, T. and Tanaka, Y. and Tamasaku, K. and Sawada, K. and Ishikawa, T.",
    title = "{Results of a Search for Paraphotons with Intense X-ray Beams at SPring-8}",
    eprint = "1301.6557",
    archivePrefix = "arXiv",
    primaryClass = "physics.ins-det",
    doi = "10.1016/j.physletb.2013.04.033",
    journal = "Phys. Lett. B",
    volume = "722",
    pages = "301--304",
    year = "2013"
}

@article{Parker:2013fxa,
    author = "Parker, Stephen R. and Hartnett, John G. and Povey, Rhys G. and Tobar, Michael E.",
    title = "{Cryogenic resonant microwave cavity searches for hidden sector photons}",
    eprint = "1410.5244",
    archivePrefix = "arXiv",
    primaryClass = "hep-ex",
    doi = "10.1103/PhysRevD.88.112004",
    journal = "Phys. Rev. D",
    volume = "88",
    pages = "112004",
    year = "2013"
}

@article{Nguyen:2019xuh,
    author = "Nguyen, Le Hoang and Lobanov, Andrei and Horns, Dieter",
    title = "{First results from the WISPDMX radio frequency cavity searches for hidden photon dark matter}",
    eprint = "1907.12449",
    archivePrefix = "arXiv",
    primaryClass = "hep-ex",
    doi = "10.1088/1475-7516/2019/10/014",
    journal = "JCAP",
    volume = "10",
    pages = "014",
    year = "2019"
}

@article{near-infrared,
  author    = {Hai-Yun Shen and Wen-Min Wang and Hong-Ling Gao and Jian-Zhong Cui},
  title     = {Near-infrared luminescence and SMM behaviors of a family of dinuclear lanthanide 8-quinolinolate complexes},
  journal   = {RSC Advances},
  year      = {2016},
  volume    = {6},
  number    = {41},
  pages     = {34165--34174},
  doi       = {10.1039/C6RA02656G},
  url       = {https://doi.org/10.1039/C6RA02656G}
}

@misc{AxionLimits,
  author       = {Ciaran O'Hare},
  title        = {cajohare/AxionLimits: AxionLimits},
  month        = jul,
  year         = 2020,
  publisher    = {Zenodo},
  version      = {v1.0},
  doi          = {10.5281/zenodo.3932430},
  howpublished = {\url{https://cajohare.github.io/AxionLimits/}}
}

@article{suzuki_mn12,
  title = {Propagation of Avalanches in ${\mathrm{Mn}}_{12}$-Acetate: Magnetic Deflagration},
  author = {Suzuki, Yoko and Sarachik, M. P. and Chudnovsky, E. M. and McHugh, S. and Gonzalez-Rubio, R. and Avraham, Nurit and Myasoedov, Y. and Zeldov, E. and Shtrikman, H. and Chakov, N. E. and Christou, G.},
  journal = {Phys. Rev. Lett.},
  volume = {95},
  issue = {14},
  pages = {147201},
  numpages = {4},
  year = {2005},
  month = {Sep},
  publisher = {American Physical Society},
  doi = {10.1103/PhysRevLett.95.147201},
  url = {https://link.aps.org/doi/10.1103/PhysRevLett.95.147201}
}

@article{magneto-chiral,
  author    = {Maria Sara Raju and Kais Dhbaibi and Maxime Grasser and Vincent Dorcet and Ivan Breslavetz and Kévin Paillot and Nicolas Vanthuyne and Olivier Cador and Geert L. J. A. Rikken and Boris Le Guennic and Jeanne Crassous and Fabrice Pointillart and Cyrille Train and Matteo Atzori},
  title     = {Magneto-Chiral Dichroism in a One-Dimensional Assembly of Helical Dysprosium(III) Single-Molecule Magnets},
  journal   = {Inorganic Chemistry},
  year      = {2023},
  volume    = {62},
  number    = {43},
  pages     = {17583--17587},
  doi       = {10.1021/acs.inorgchem.3c03204},
  url       = {https://doi.org/10.1021/acs.inorgchem.3c03204}
}

@article{An_2015,
   title={Direct detection constraints on dark photon dark matter},
   volume={747},
   ISSN={0370-2693},
   url={http://dx.doi.org/10.1016/j.physletb.2015.06.018},
   DOI={10.1016/j.physletb.2015.06.018},
   journal={Physics Letters B},
   publisher={Elsevier BV},
   author={An, Haipeng and Pospelov, Maxim and Pradler, Josef and Ritz, Adam},
   year={2015},
   month=jul, pages={331–338} }

@article{Miyazaki2001,
  author    = {Yuji Miyazaki and Ashis Bhattacharjee and Motohiro Nakano and Kazuya Saito and Sheila M. J. Aubin and Hilary J. Eppley and George Christou and David N. Hendrickson and Michio Sorai},
  title     = {Magnetic-Field-Dependent Heat Capacity of the Single-Molecule Magnet [{Mn12O12(O2CEt)16(H2O)3}]},
  journal   = {Inorganic Chemistry},
  year      = {2001},
  volume    = {40},
  number    = {26},
  pages     = {6632--6636},
  doi       = {10.1021/ic010567w},
  url       = {https://doi.org/10.1021/ic010567w},
  publisher = {American Chemical Society}
}

@article{Bringmann:2018cvk,
    author = "Bringmann, Torsten and Pospelov, Maxim",
    title = "{Novel direct detection constraints on light dark matter}",
    eprint = "1810.10543",
    archivePrefix = "arXiv",
    primaryClass = "hep-ph",
    doi = "10.1103/PhysRevLett.122.171801",
    journal = "Phys. Rev. Lett.",
    volume = "122",
    number = "17",
    pages = "171801",
    year = "2019"
}

@article{Caputo_2021,
   title={Dark photon limits: A handbook},
   volume={104},
   ISSN={2470-0029},
   url={http://dx.doi.org/10.1103/PhysRevD.104.095029},
   DOI={10.1103/physrevd.104.095029},
   number={9},
   journal={Physical Review D},
   publisher={American Physical Society (APS)},
   author={Caputo, Andrea and Millar, Alexander J. and O’Hare, Ciaran A. J. and Vitagliano, Edoardo},
   year={2021},
   month=nov }

@article{Friedman_2010,
   title={Single-Molecule Nanomagnets},
   volume={1},
   ISSN={1947-5462},
   url={http://dx.doi.org/10.1146/annurev-conmatphys-070909-104053},
   DOI={10.1146/annurev-conmatphys-070909-104053},
   number={1},
   journal={Annual Review of Condensed Matter Physics},
   publisher={Annual Reviews},
   author={Friedman, Jonathan R. and Sarachik, Myriam P.},
   year={2010},
   month=aug, pages={109–128} }

@article{Briganti2021,
  author    = {Matteo Briganti and Fabio Santanni and Lorenzo Tesi and Federico Totti and Roberta Sessoli and Alessandro Lunghi},
  title     = {A Complete Ab Initio View of Orbach and Raman Spin–Lattice Relaxation in a Dysprosium Coordination Compound},
  journal   = {J. Am. Chem. Soc.},
  year      = {2021},
  volume    = {143},
  number    = {34},
  pages     = {13633--13645},
  doi       = {10.1021/jacs.1c05068},
  url       = {https://doi.org/10.1021/jacs.1c05068}
}

@article{XENON:2023cxc,
    author = "Aprile, E. and others",
    collaboration = "XENON",
    title = "{First Dark Matter Search with Nuclear Recoils from the XENONnT Experiment}",
    eprint = "2303.14729",
    archivePrefix = "arXiv",
    primaryClass = "hep-ex",
    doi = "10.1103/PhysRevLett.131.041003",
    journal = "Phys. Rev. Lett.",
    volume = "131",
    number = "4",
    pages = "041003",
    year = "2023"
}

@article{Fermi-LAT:2016uux,
    author = "Albert, A. and others",
    collaboration = "Fermi-LAT, DES",
    title = "{Searching for Dark Matter Annihilation in Recently Discovered Milky Way Satellites with Fermi-LAT}",
    eprint = "1611.03184",
    archivePrefix = "arXiv",
    primaryClass = "astro-ph.HE",
    reportNumber = "FERMILAB-PUB-16-073-AE",
    doi = "10.3847/1538-4357/834/2/110",
    journal = "Astrophys. J.",
    volume = "834",
    number = "2",
    pages = "110",
    year = "2017"
}

@article{XENON:2024znc,
    author = "Aprile, E. and others",
    collaboration = "XENON",
    title = "{Search for Light Dark Matter in Low-Energy Ionization Signals from XENONnT}",
    eprint = "2411.15289",
    archivePrefix = "arXiv",
    primaryClass = "hep-ex",
    doi = "10.1103/PhysRevLett.134.161004",
    journal = "Phys. Rev. Lett.",
    volume = "134",
    number = "16",
    pages = "161004",
    year = "2025"
}

@inproceedings{PerezAdan:2023rsl,
    author = "Perez Adan, Danyer",
    collaboration = "ATLAS, CMS",
    title = "{Dark Matter searches at CMS and ATLAS}",
    booktitle = "{56th Rencontres de Moriond on Electroweak Interactions and Unified Theories}",
    eprint = "2301.10141",
    archivePrefix = "arXiv",
    primaryClass = "hep-ex",
    reportNumber = "CMS-CR-2022-059",
    month = "1",
    year = "2023"
}

@article{Kahn:2021ttr,
    author = "Kahn, Yonatan and Lin, Tongyan",
    title = "{Searches for light dark matter using condensed matter systems}",
    eprint = "2108.03239",
    archivePrefix = "arXiv",
    primaryClass = "hep-ph",
    doi = "10.1088/1361-6633/ac5f63",
    journal = "Rept. Prog. Phys.",
    volume = "85",
    number = "6",
    pages = "066901",
    year = "2022"
}

@article{Hochberg:2015fth,
    author = "Hochberg, Yonit and Pyle, Matt and Zhao, Yue and Zurek, Kathryn M.",
    title = "{Detecting Superlight Dark Matter with Fermi-Degenerate Materials}",
    eprint = "1512.04533",
    archivePrefix = "arXiv",
    primaryClass = "hep-ph",
    doi = "10.1007/JHEP08(2016)057",
    journal = "JHEP",
    volume = "08",
    pages = "057",
    year = "2016"
}

@article{Hochberg_2016,
   title={Detecting superlight dark matter with Fermi-degenerate materials},
   volume={2016},
   ISSN={1029-8479},
   url={http://dx.doi.org/10.1007/JHEP08(2016)057},
   DOI={10.1007/jhep08(2016)057},
   number={8},
   journal={Journal of High Energy Physics},
   publisher={Springer Science and Business Media LLC},
   author={Hochberg, Yonit and Pyle, Matt and Zhao, Yue and Zurek, Kathryn M.},
   year={2016},
   month=aug }

@article{Hochberg:2016sqx,
    author = "Hochberg, Yonit and Lin, Tongyan and Zurek, Kathryn M.",
    title = "{Absorption of light dark matter in semiconductors}",
    eprint = "1608.01994",
    archivePrefix = "arXiv",
    primaryClass = "hep-ph",
    doi = "10.1103/PhysRevD.95.023013",
    journal = "Phys. Rev. D",
    volume = "95",
    number = "2",
    pages = "023013",
    year = "2017"
}

@article{Super-Kamiokande:2022ncz,
    author = "Abe, K. and others",
    collaboration = "Super-Kamiokande",
    title = "{Search for Cosmic-Ray Boosted Sub-GeV Dark Matter Using Recoil Protons at Super-Kamiokande}",
    eprint = "2209.14968",
    archivePrefix = "arXiv",
    primaryClass = "hep-ex",
    doi = "10.1103/PhysRevLett.130.031802",
    journal = "Phys. Rev. Lett.",
    volume = "130",
    number = "3",
    pages = "031802",
    year = "2023",
    note = "[Erratum: Phys.Rev.Lett. 131, 159903 (2023)]"
}

@article{Das:2025qxm,
    author = "Das, Anirban and Herbermann, Tim and Sen, Manibrata and Takhistov, Volodymyr",
    title = "{Energy-dependent boosted DM from DSNB}",
    doi = "10.22323/1.473.0014",
    journal = "PoS",
    volume = "NOW2024",
    pages = "014",
    year = "2025"
}

@article{Herbermann:2024kcy,
    author = "Herbermann, Tim and Lindner, Manfred and Sen, Manibrata",
    title = "{Attenuation of cosmic ray electron boosted dark matter}",
    eprint = "2408.02721",
    archivePrefix = "arXiv",
    primaryClass = "hep-ph",
    doi = "10.1103/PhysRevD.110.123023",
    journal = "Phys. Rev. D",
    volume = "110",
    number = "12",
    pages = "123023",
    year = "2024"
}

@article{Das:2024ghw,
    author = "Das, Anirban and Herbermann, Tim and Sen, Manibrata and Takhistov, Volodymyr",
    title = "{Energy-dependent boosted dark matter from diffuse supernova neutrino background}",
    eprint = "2403.15367",
    archivePrefix = "arXiv",
    primaryClass = "hep-ph",
    doi = "10.1088/1475-7516/2024/07/045",
    journal = "JCAP",
    volume = "07",
    pages = "045",
    year = "2024"
}

@article{ParticleDataGroup:2024cfk,
    author = "Navas, S. and others",
    collaboration = "Particle Data Group",
    title = "{Review of particle physics}",
    doi = "10.1103/PhysRevD.110.030001",
    journal = "Phys. Rev. D",
    volume = "110",
    number = "3",
    pages = "030001",
    year = "2024"
}

@article{axion-absorption,
    author = "Berlin, Asher and Trickle, Tanner",
    title = "{Absorption of Axion Dark Matter in a Magnetized Medium}",
    eprint = "2305.05681",
    archivePrefix = "arXiv",
    primaryClass = "hep-ph",
    reportNumber = "FERMILAB-PUB-23-168-SQMS-T",
    doi = "10.1103/PhysRevLett.132.181801",
    journal = "Phys. Rev. Lett.",
    volume = "132",
    number = "18",
    pages = "181801",
    year = "2024"
}

@article{Mitridate:2021ctr,
    author = "Mitridate, Andrea and Trickle, Tanner and Zhang, Zhengkang and Zurek, Kathryn M.",
    title = "{Dark matter absorption via electronic excitations}",
    eprint = "2106.12586",
    archivePrefix = "arXiv",
    primaryClass = "hep-ph",
    reportNumber = "CALT-TH-2021-025",
    doi = "10.1007/JHEP09(2021)123",
    journal = "JHEP",
    volume = "09",
    pages = "123",
    year = "2021"
}

@article{Catena:2022fnk,
    author = "Catena, Riccardo and Cole, Daniel and Emken, Timon and Matas, Marek and Spaldin, Nicola and Tarantino, Walter and Urdshals, Einar",
    title = "{Dark matter-electron interactions in materials beyond the dark photon model}",
    eprint = "2210.07305",
    archivePrefix = "arXiv",
    primaryClass = "hep-ph",
    doi = "10.1088/1475-7516/2023/03/052",
    journal = "JCAP",
    volume = "03",
    pages = "052",
    year = "2023"
}

@article{Boveia:2018yeb,
    author = "Boveia, Antonio and Doglioni, Caterina",
    title = "{Dark Matter Searches at Colliders}",
    eprint = "1810.12238",
    archivePrefix = "arXiv",
    primaryClass = "hep-ex",
    doi = "10.1146/annurev-nucl-101917-021008",
    journal = "Ann. Rev. Nucl. Part. Sci.",
    volume = "68",
    pages = "429--459",
    year = "2018"
}

@article{Granelli:2022ysi,
    author = "Granelli, Alessandro and Ullio, Piero and Wang, Jin-Wei",
    title = "{Blazar-boosted dark matter at Super-Kamiokande}",
    eprint = "2202.07598",
    archivePrefix = "arXiv",
    primaryClass = "astro-ph.HE",
    reportNumber = "SISSA 02/2022/FISI",
    doi = "10.1088/1475-7516/2022/07/013",
    journal = "JCAP",
    volume = "07",
    number = "07",
    pages = "013",
    year = "2022"
}

@article{Gaskins:2016cha,
    author = "Gaskins, Jennifer M.",
    title = "{A review of indirect searches for particle dark matter}",
    eprint = "1604.00014",
    archivePrefix = "arXiv",
    primaryClass = "astro-ph.HE",
    doi = "10.1080/00107514.2016.1175160",
    journal = "Contemp. Phys.",
    volume = "57",
    number = "4",
    pages = "496--525",
    year = "2016"
}

@article{PropagationofAvalanches,
  title = {Propagation of Avalanches in ${\mathrm{Mn}}_{12}$-Acetate: Magnetic Deflagration},
  author = {Suzuki, Yoko and Sarachik, M. P. and Chudnovsky, E. M. and McHugh, S. and Gonzalez-Rubio, R. and Avraham, Nurit and Myasoedov, Y. and Zeldov, E. and Shtrikman, H. and Chakov, N. E. and Christou, G.},
  journal = {Phys. Rev. Lett.},
  volume = {95},
  issue = {14},
  pages = {147201},
  numpages = {4},
  year = {2005},
  month = {Sep},
  publisher = {American Physical Society},
  doi = {10.1103/PhysRevLett.95.147201},
  url = {https://link.aps.org/doi/10.1103/PhysRevLett.95.147201}
}

@article{Orts,
author = {Orts-Arroyo, Marta and Rojas, Carlos and Moliner, Nicolás and Martínez-Lillo, José},
year = {2023},
month = {05},
pages = {8645},
title = {Lipoic Acid-Functionalized Hexanuclear Manganese(III) Nanomagnets Suitable for Surface Grafting},
volume = {24},
journal = {International Journal of Molecular Sciences},
doi = {10.3390/ijms24108645}
}

@article{ZABALALEKUONA2021213984,
title = {Single-Molecule Magnets: From Mn12-ac to dysprosium metallocenes, a travel in time},
journal = {Coordination Chemistry Reviews},
volume = {441},
pages = {213984},
year = {2021},
issn = {0010-8545},
doi = {https://doi.org/10.1016/j.ccr.2021.213984},
url = {https://www.sciencedirect.com/science/article/pii/S0010854521002587},
author = {Andoni Zabala-Lekuona and José Manuel Seco and Enrique Colacio}
}

@article{Sessoli,
author = {Sessoli, Roberta and Gatteschi, Dante and Caneschi, Andrea and Novak, Miguel},
year = {1993},
month = {09},
pages = {141-143},
title = {Magnetic bistability in a metal-ion cluster},
volume = {365},
journal = {Nature},
doi = {10.1038/365141a0}
}

@article{Kim:2008hd,
    author = "Kim, Jihn E. and Carosi, Gianpaolo",
    title = "{Axions and the Strong CP Problem}",
    eprint = "0807.3125",
    archivePrefix = "arXiv",
    primaryClass = "hep-ph",
    doi = "10.1103/RevModPhys.82.557",
    journal = "Rev. Mod. Phys.",
    volume = "82",
    pages = "557--602",
    year = "2010",
    note = "[Erratum: Rev.Mod.Phys. 91, 049902 (2019)]"
}

@article{Preskill:1982cy,
    author = "Preskill, John and Wise, Mark B. and Wilczek, Frank",
    editor = "Srednicki, M. A.",
    title = "{Cosmology of the Invisible Axion}",
    reportNumber = "HUTP-82-A048, NSF-ITP-82-103",
    doi = "10.1016/0370-2693(83)90637-8",
    journal = "Phys. Lett. B",
    volume = "120",
    pages = "127--132",
    year = "1983"
}

@article{Peccei:1977hh,
    author = "Peccei, R. D. and Quinn, Helen R.",
    title = "{CP Conservation in the Presence of Instantons}",
    reportNumber = "ITP-568-STANFORD",
    doi = "10.1103/PhysRevLett.38.1440",
    journal = "Phys. Rev. Lett.",
    volume = "38",
    pages = "1440--1443",
    year = "1977"
}

@Article{C9SC01062A,
author ="Moreno-Pineda, Eufemio and Taran, Gheorghe and Wernsdorfer, Wolfgang and Ruben, Mario",
title  ="Quantum tunnelling of the magnetisation in single-molecule magnet isotopologue dimers",
journal  ="Chem. Sci.",
year  ="2019",
volume  ="10",
issue  ="19",
pages  ="5138-5145",
publisher  ="The Royal Society of Chemistry",
doi  ="10.1039/C9SC01062A",
url  ="http://dx.doi.org/10.1039/C9SC01062A"
}

@book{abragam2012electron,
  title={Electron Paramagnetic Resonance of Transition Ions},
  author={Abragam, A. and Bleaney, B.},
  isbn={9780191023002},
  series={Oxford Classic Texts in the Physical Sciences},
  url={https://books.google.com.br/books?id=ASNoAgAAQBAJ},
  year={2012},
  publisher={Oxford}
}

@ARTICLE{2001SoPh,
       author = {{Charbonneau}, Paul and {McIntosh}, Scott W. and {Liu}, Han-Li and {Bogdan}, Thomas J.},
        title = "{Avalanche models for solar flares (Invited Review)}",
      journal = {Solphys},
         year = 2001,
        month = nov,
       volume = {203},
       number = {2},
        pages = {321-353},
          doi = {10.1023/A:1013301521745},
       adsurl = {https://ui.adsabs.harvard.edu/abs/2001SoPh..203..321C}
}

@article{McLuckie_1988,
doi = {10.1088/0031-9120/23/4/417},
url = {https://doi.org/10.1088/0031-9120/23/4/417},
year = {1988},
month = {jul},
publisher = {},
volume = {23},
number = {4},
pages = {251},
author = {I F McLuckie},
title = {Investigation of Zener and avalanche diode characteristics},
journal = {Physics Education}
}

\end{document}